\documentclass[useAMS,usenatbib]{mn2e}
\usepackage{graphicx}

\newcommand{\ap}{\approx}

\newcommand{\kms}{km~s$^{-1}$}

\newcommand{\hi}{H\,{\sc i}}
\newcommand{\hii}{H\,{\sc ii}}

\newcommand{\emax}{\ensuremath{\epsilon_{\mathrm{max}}}}

\newcommand{\amax}{\ensuremath{a_{\epsilon}}}
\newcommand{\amin}{\ensuremath{a_{\mathrm{min}}}}
\newcommand{\aten}{\ensuremath{a_{10}}}
\newcommand{\lbar}{\ensuremath{L_{\mathrm{bar}}}}
\newcommand{\lavg}{\ensuremath{L_{\mathrm{avg}}}}
\newcommand{\lph}{\ensuremath{L_{\mathrm{phase}}}}

\title[Bar Sizes]{How Large Are the Bars in Barred Galaxies?}

\author[Peter Erwin]{Peter Erwin$^{1,2,3}$\thanks{E-mail: 
erwin@mpe.mpg.de}\\
$^{1}$Instituto de Astrofisica de Canarias, C/ V\'{\i}a L\'{a}ctea s/n, 
38200 La Laguna, Tenerife, Spain\\
$^{2}$Current address: Max-Planck-Institut f\"{u}r extraterrestrische 
Physik, Giessenbachstrasse, D-85748 Garching, Germany\\
$^{3}$Guest investigator of the UK Astronomy Data Centre}

\begin{document}



\maketitle

\label{firstpage}

\begin{abstract} 
I present a study of the sizes (semimajor axes) of bars in disc
galaxies, combining a detailed $R$-band study of 65 S0--Sb galaxies
with the $B$-band measurements of 70 Sb--Sd galaxies from Martin
(1995).  As has been noted before with smaller samples, bars in
early-type (S0--Sb) galaxies are clearly larger than bars in late-type
(Sc--Sd) galaxies; this is true both for relative sizes (bar length as
fraction of isophotal radius $R_{25}$ or exponential disc scale length
$h$) and absolute sizes (kpc).  S0--Sab bars extend to $\sim1$--10 kpc
(mean $\sim3.3$ kpc), $\sim0.2$--0.8 $R_{25}$ (mean $\sim0.38 \;
R_{25}$) and $\sim 0.5$--2.5 $h$ (mean $\sim 1.4 \; h$).  Late-type
bars extend to only $\sim0.5$--3.5 kpc, $\sim0.05$--0.35 $R_{25}$ and
0.2--1.5 $h$; their mean sizes are $\sim 1.5$ kpc, $\sim0.14 \;
R_{25}$ and $\sim0.6 \; h$.  Sb galaxies resemble earlier-type
galaxies in terms of bar size relative to $h$; their smaller
$R_{25}$-relative sizes may be a side effect of higher star formation,
which increases $R_{25}$ but not $h$.  Sbc galaxies form a transition
between the early- and late-type regimes.  For S0--Sbc galaxies, bar
size correlates well with disc size (both $R_{25}$ and $h$); these
correlations are stronger than the known correlation with $M_{B}$.
All correlations appear to be weaker or absent for late-type galaxies;
in particular, there seems to be \textit{no} correlation between bar
size and either $h$ or $M_{B}$ for Sc--Sd galaxies.

Since bar size scales with disc size and galaxy magnitude for most
Hubble types, studies of bar evolution with redshift should select
samples with similar distributions of disc size or magnitude
(extrapolated to present-day values); otherwise, bar frequencies and
sizes could be mis-estimated.  Because early-type galaxies tend to
have larger bars, resolution-limited studies will preferentially find
bars in early-type galaxies (assuming no significant differential
evolution in bar sizes).  I show that the bars detected in
\textit{HST} near-IR images at $z \sim 1$ by Sheth et al.\ (2003) have
absolute sizes consistent with those in bright, nearby S0--Sb
galaxies.  I also compare the sizes of real bars with those produced
in simulations, and discuss some possible implications for scenarios
of secular evolution along the Hubble sequence.  Simulations often
produce bars as large as -- or larger than -- those seen in S0--Sb
galaxies, but rarely any as small as those in Sc--Sd galaxies.

\end{abstract}


\section{Introduction} 

Observations indicate that $\sim70$\% of all disc galaxies are barred
to one degree or another \citep[e.g.,][]{eskridge00,erwin05-rc3}.
There is considerable debate about the origin and influence of bars,
and also about their strengths, something for which there is still no
agreed-upon measurement, though many have been suggested
\citep[e.g.,][]{m95,seigar98,chapelon99,buta01}.  Curiously, somewhat
less attention has been given to the question of bar \textit{sizes},
perhaps because this seems, on the face of it, easier to measure --
even though there are no agreed-upon methods of measuring bar sizes
either (see the discussions in \nocite{athan-m02}Athanassoula \&
Misiriotis 2002 and \nocite{aguerri03}Aguerri et al.\ 2003).

Is bar size actually important?  There are, I would argue, several
reasons why bar size is interesting, beyond a basic natural
historian's curiosity (``Just how big or small \textit{are} they,
anyway?'').  To begin with, the size of a bar is first approximation
to how much of its host galaxy can be affected by the bars' dynamical
influence: larger bars can obviously affect more of the galaxy than
smaller bars.  In addition to the well-known effects of bars on gas
flow, \citet{weinberg02} and \citet{holley-b05} argued that bars can
also restructure dark-matter halos, flattening out the steep central
cusps which are produced in cosmological simulations but apparently
not seen in real galaxies \citep[but see][]{sellwood03,athan04}.
Larger bars could then mean larger dark-matter cores.
Holley-Bockelmann et al.\ also argued that tidally-induced bars could
be significantly larger than the ``classical'' $n$-body bars which
form via disc instabilities, so bar size may provide clues to past
merger histories.  More generally, bar size is an obvious way of
testing the accuracy of different bar-formation and bar-evolution
models \citep[e.g.,][]{vk03}.  For example, \citet{bournaud02}
recently outlined a scenario of galaxy evolution involving multiple
rounds of bar formation, self-destruction, and resurrection due to gas
accretion; they predict a trend of bar size with Hubble type, where
galaxies with larger bulges (i.e., earlier Hubble types) have shorter
bars.  $N$-body simulations also suggest that bar size depends on
angular momentum exchange between the bar and the bulge, the outer
disc, and the halo.  Since the relative masses of these components, as
well as how kinematically hot they are, can affect how much angular
mometum is exchanged \citep[e.g.,][]{athan03}, bar size could be a
useful probe of halo mass and kinematics.  Finally, several studies
have suggested that longer bars are correlated with higher star
formation activity, at least in late-type galaxies
\citep[e.g.,][]{martinet97,chapelon99}.

The first systematic investigation of bar sizes was made by
\citet{kormendy79}, who found that bar size correlated with galaxy
blue luminosity.  Subsequently, \citet[][ hereafter EE85]{ee85} showed
that bars in early-type disc galaxies tended to be larger, relative to
the optical disc diameter $D_{25}$, than bars in later Hubble types
\citep[see also][]{regan97}.  They also found a dichotomy in bar
\textit{structure}: early-type galaxies typically have bars with
shallow (``flat'') profiles and truncations \citep[as noted
by][]{kormendy82}, while late-type galaxies tend to have bars with
steep exponential profiles.  More recent studies using CCDs or
near-infrared images include those of \citet{chapelon99},
\citet{laine02}, and Laurikainen and collaborators \citep{ls02,lsr02};
these have, in general, supported the findings of Kormendy and EE85.

Because a number of these studies have focused on particular subtypes
of galaxies, the results are not as general or unbiased as they might
otherwise be.  For example, \citet{chapelon99} studied primarily a
large sample of starburst galaxies, while the studies of
\citet{laine02} and \citet{lsr02} were aimed at Seyfert and other
``active'' galaxies.  Chapelon et al.\ noted that bar sizes for their
late-type (starburst) galaxies tended to be larger than those of the
normal late-type galaxies of \citet{m95}; similarly, Laurikainen et
al.\ found that Seyferts tended to have larger bars than non-Seyferts.
Except for the pioneering studies of Kormendy and EE85, bars in the
earliest disc galaxies -- i.e., S0 galaxies -- have been 
ignored or represented by only a handful of examples.  Finally, there
has also been a tendency to overlook so-called ``weak'' (i.e., SAB)
bars: the samples of Kormendy and EE85 are almost entirely SB
galaxies, and bar sizes were measured by Laurikainen et al.\ only for
galaxies with relatively high $m = 2$ Fourier bar amplitudes.

Thus, there is still considerable room for improving our understanding
of bar sizes in the general population of disc galaxies, especially
for S0 galaxies and weak bars.  The relevance of bar size
distributions for galaxy evolution was recently highlighted by
\citet{sheth03}, who discussed the visibility of bars as a function of
redshift and resolution.  Put simply, large bars (size in kpc) are
easier to detect at high redshift than small bars; if the average bar
is small enough, it will be undetectable at high $z$.  Failure to
account for this possibility can produce spurious changes in bar
fraction with redshift.  (We might also like to know if the average 
bar size has changed significantly between, say, $z = 1$ and now, 
which presupposes a good understanding of local bar sizes.)

In this paper, I take a detailed look at the question of bar sizes
along the Hubble sequence in the local universe.  The main part of
this study uses a diameter-limited sample of nearby, early-type
(S0--Sb) disc galaxies with both strong (SB) and weak (SAB) bars.  I
measure the bar lengths and compare them with the overall size of the
galaxy, using both the 25th-magnitude radius ($R_{25}$) and the
exponential scale lengths of the outer discs.  These are combined
with the measurements of \citet{m95}, which also include both SB and
SAB bars and are primarily of later Hubble types (Sbc--Sd).

\section{Samples} 

I use two samples of galaxies in this paper.  The first is a sample of
early-type (S0--Sb) galaxies, using recent $R$-band imaging for both
bar-size and exponential disc scale length measurements; both sets of
measurements are presented here.  To extend this study to later Hubble
types, I have drawn on a second sample, that of \citet{m95}.  This
consists primarily of Sb--Sd galaxies, with bar-size measurements made
from blue photographic prints.

The early-type galaxy sample is an expanded version of that presented
in \citet{erwin-sparke03}; I will refer to their original sample as
the ``WIYN Sample'' (since most of the observations were made with the
3.5m WIYN Telescope).\footnote{The WIYN Observatory is a joint
facility of the University of Wisconsin-Madison, Indiana University,
Yale University, and the National Optical Astronomy Observatories} The
WIYN Sample consists of all optically barred (SB + SAB) S0--Sa
galaxies from the UGC catalog \citep{ugc} which met the following
criteria: declination $> 10\degr$, heliocentric radial velocity $\leq
2000$ \kms, major axis diameter $\ge 2\arcmin$, and ratio of major to
minor axis $a/b \leq 2$ (corresponding to $i \la 60\degr$).  Galaxy
types and axis measurements (at the 25 mag arcsec$^{-2}$ level in $B$)
were taken from \citet[][hereafter RC3]{rc3}; radial velocities are
from the NASA/IPAC Extragalactic Database (NED).  The size restriction
and the use of the UGC means that the sample is biased in favor of
bright, high surface brightness galaxies.  The sample had a total of
38 galaxies; I subsequently eliminated four galaxies where bars were
either absent, ambiguous, or too difficult to measure (see the
Appendix), leaving a total of 34 S0--Sa galaxies.

There is some evidence that Hubble types in clusters -- in the Virgo
Cluster, at least -- do not agree with Hubble types of isolated field
galaxies \citep{vandb76}.  \citet{koopman98} found that Virgo Sa--Sab
galaxies had central light concentrations more like those of isolated
Sb--Sc field galaxies.  Accordingly, Erwin \& Sparke excluded Virgo
galaxies from their sample.  Because the case for S0 galaxies is
unclear (Koopman \& Kenney noted that their sample was strongly
incomplete for S0 galaxies, and the few S0 galaxies they studied did
not differ significantly between field and Virgo -- see their
Figure~1), and because information on bar sizes in S0 galaxies is
particularly lacking, I have added bar measurements for eight of the
ten barred S0's in Virgo which meet the criteria given above, except
for the redshift limit.  (The redshift limit was intended to set a
distance limit of $\sim 30$ Mpc for the field galaxies; if it were
applied to the Virgo Cluster, which lies well within that distance, it
would improperly exclude cluster galaxies with high peculiar
velocities.)

To make the coverage of Hubble types more complete, I have also added
galaxies from an ongoing study of barred Sab and Sb galaxies (Erwin,
Vega Beltr\'an \& Beckman, in preparation).  The selection criteria
are identical, aside from the difference in Hubble type, leading to a
total of 9 Sab and 18 Sb galaxies; two of each type appear to be
unbarred, and are not considered further (see the Appendix).

The final early-type sample thus has a total of 65 strongly and weakly
barred S0--Sb galaxies.  All of these galaxies, grouped by Hubble
type, are listed in Table~\ref{tab:galaxies}, along with the
parameters of their bars and outer discs.

\begin{figure}  
\includegraphics[scale=0.5]{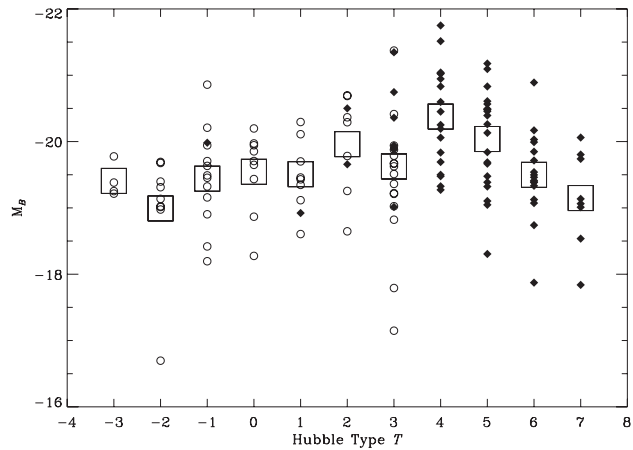}
\caption{Absolute blue magnitudes for galaxies from the two samples as
a function of Hubble type.  Galaxies from my sample are shown with
open circles, while galaxies from \citet{m95} meeting the same
selection criteria are shown with filled diamonds; mean values for
each Hubble type are indicated by the large boxes.}
\label{fig:absmags}
\end{figure}

The sample which best complements mine is that of \citet{m95}: it is
large and drawn from ordinary, optically barred galaxies (including
both SB and SAB classes), contains both observed and deprojected bar
lengths, and is almost entirely Sb or later in Hubble type.  To make
the match between samples as close as possible, I applied the same
selection criteria to Martin's galaxies: SB or SAB bar classification,
major axis $\geq 2\arcmin$, axis ratio $\leq 2$, and radial velocity
$\leq 2000$ \kms.  Martin argued that deprojection was unreliable for
Magellanic galaxies (Sm and Im), so I follow him in excluding those
types.  I also removed three Virgo galaxies (NGC~4303, NGC~4321, and
NGC~4394) and eliminated NGC~4395, which is classed as SA in
\nocite{rc3}RC3; for the three galaxies in common between the samples
(see below), I retain my measurements.  This leaves a total of 75
galaxies from his sample, still large enough for a good comparison;
the bulk of these (70) are Sb--Sd.  To these I added distances and
total blue magnitudes, mostly from LEDA (see Appendix~\ref{app:m95}
for details).  Although the underlying sample selection was different
(Martin's galaxies were taken from the Sandage-Bedke atlas), the
relative numbers of different Hubble types are consistent with local
populations.  For example, my sample has 24 S0/a--Sab galaxies
compared with 56 Sbc--Scd galaxies in Martin's sample; the ratio of
late to early types (2.3) is similar to that found in RC3 for galaxies
with $D_{25} \geq 2.0\arcmin$ and $a/b \leq 2.0$ (480 Sbc--Scd
galaxies versus 199 S0/a--Sab, for a ratio of 2.4).  This suggests
that the combined set provides a reasonable picture of bar sizes for
the Hubble sequence down to Sd (Martin's sample has very few Sdm or Sm
galaxies), at least for bright galaxies (median $M_{B} = -19.5$ for my
S0--Sb galaxies and $-19.8$ for Martin's Sb--Sd; see
Figure~\ref{fig:absmags}).

For the early-type galaxies, the measurements of bar size and shape,
and of disc sizes, are discussed in
Sections~\ref{sec:measure-bars}--\ref{sec:measure-discs}, below.  In
Section~\ref{sec:m95}, I discuss how the published bar sizes of
\citet{m95} can best be compared with my measurements, and how I
obtained disc scale lengths for Martin's galaxies.

\section{Observations and Measurements} 
\label{sec:obs}

Observations of the original WIYN Sample (S0--Sa galaxies) are
presented and discussed in detail by \citet{erwin-sparke03}.  All but
two of the galaxies were observed in $B$ and $R$ with the 3.5 m WIYN
Telescope at Kitt Peak, Arizona, between 1995 December and 1998 March.
Images for NGC~936 and NGC~4314 were taken from the BARS Project
observations \citep{lourenso01}; in a few cases, additional images
from other sources were used for outer-disc or bar measurements (see
the Appendix for details).

Images of the barred Virgo S0 galaxies and the Sab--Sb galaxies are
from a variety of sources, including the WIYN Telescope (1995 March
through 1998 March); the 2.4 m MDM Telescope at Kitt Peak, courtesy
Paul Schechter (1996 March); the 2.5 m Nordic Optical Telescope in La
Palma (2001 April and 2002 April); and the Isaac Newton Group archive
(images from both the 1 m Jacobus Kapteyn Telescope and the 2.5 m
Isaac Newton Telescope).  The outer-disc scale lengths for several
galaxies were measured using images taken with the Isaac Newton
Telescope in 2004 March; these observations will be described in more
detail in Erwin, Pohlen, \& Beckman (in preparation).  I also used
observations from the BARS Project (for NGC~4151 and NGC~4596) and
$r$- or $R$-band images from the sample of \citet{frei96} for a number
of galaxies.  Specific details for individual galaxies are discussed
in the Appendix.

Except where noted in the Appendix (cases where dust severely distorted
bar isophotes in the optical images), all bar and outer-disc scale
length measurements were made with $R$-band or equivalent images.  For
Sab and Sb galaxies, these measurements were usually checked against
measurements made with near-IR images; the agreement was generally
very good.

Parameters for these galaxies are listed in Table~\ref{tab:galaxies}.
Most distances are from LEDA (exceptions are discussed in the
Appendix); the latter are based on redshifts corrected for
Virgocentric infall, as listed in LEDA, and assuming $H_{0} = 75$
\kms{} kpc$^{-1}$.  For Virgo galaxies, I assumed a default distance
to the Virgo Cluster of 15.3 Mpc \citep{freedman01}, except for
NGC~4754, for which a surface-brightness fluctuation measurement by
\citet{tonry01} is available.  Note that distance measurements, and
their uncertainties, only affect the \textit{absolute} sizes of bars;
relative bar sizes are distance-independent.

\begin{table*}
\begin{minipage}{126mm}
    \caption{Bar and Disc Measurements for S0--Sb Galaxies}
    \label{tab:galaxies}
    \begin{tabular}{llrrrcrrrrrrr}
\hline
Name & Type & Distance & $M_{B}$ &  \multicolumn{4}{l}{Outer Disc}
     & \multicolumn{5}{l}{Bar} \\
     &      &          &         &  PA & $i$ & $R_{25}$ & $h$~
     & PA & \amax{} & \amin/\aten{} & \lbar{} & \emax{} \\
     &      & (Mpc)    &          & ($\degr$) & ($\degr$)
     & ($\arcsec$) & ($\arcsec$) & ($\degr$) & ($\arcsec$) 
     & ($\arcsec$)~~~ & ($\arcsec$) & \\
\hline
\multicolumn{13}{c}{S0 Galaxies} \\
 NGC 936 &            SB(rs)$0^{+}$ &  23.0 & $-20.86$ &  130 &   41 &  140 & \ldots &   81 &   41 &  65/ 51 &   51 &  0.47 \\
NGC 2787 &             SB(r)$0^{+}$ &   7.5 & $-18.20$ &  109 &   55 &   95 &     27 &  160 &   29 &  36/ 36 &   36 &  0.34 \\
NGC 2859 &          (R)SB(r)$0^{+}$ &  24.2 & $-20.21$ &   90 &   25 &  128 & \ldots &  162 &   34 &  52/ 43 &   43 &  0.40 \\
NGC 2880 &                SB$0^{-}$ &  21.9 & $-19.38$ &  144 &   52 &   62 & \ldots &   82 &    8 &   9/ 10 &    9 &  0.20 \\
NGC 2950 &          (R)SB(r)$0^{0}$ &  14.9 & $-19.14$ &  120 &   48 &   80 &     32 &  162 &   24 &  41/ 31 &   31 &  0.43 \\
NGC 2962 &        (R)SAB(rs)$0^{+}$ &  30.0 & $-19.71$ &    7 &   53 &   79 & \ldots &  168 &   29 &  43/\ldots &   43 &  0.30 \\
NGC 3412 &             SB(s)$0^{0}$ &  11.3 & $-18.98$ &  152 &   58 &  109 & \ldots &  100 &   15 &  21/ 21 &   21 &  0.26 \\
NGC 3489 &           SAB(rs)$0^{+}$ &  12.1 & $-19.45$ &   71 &   58 &  106 &     17 &   12 &   13 & \ldots/ 17 &   17 &  0.17 \\
NGC 3941 &             SB(s)$0^{0}$ &  12.2 & $-19.31$ &    9 &   52 &  104 &     25 &  166 &   21 &  36/ 32 &   32 &  0.47 \\
NGC 3945 &         (R)SB(rs)$0^{+}$ &  19.8 & $-19.94$ &  158 &   55 &  158 & \ldots &   72 &   32 &  41/ 39 &   39 &  0.29 \\
NGC 4143 &            SAB(s)$0^{0}$ &  15.9 & $-19.40$ &  144 &   59 &   68 &     14 &  163 &   17 & \ldots/ 28 &   28 &  0.38 \\
NGC 4203 &               SAB$0^{-}$ &  15.1 & $-19.21$ &   10 &   34 &  102 & \ldots &    9 &   13 &  46/\ldots &   46 &  0.24 \\
NGC 4386 &              SAB$0^{0}$: &  27.0 & $-19.68$ &  140 &   48 &   74 &     25 &  134 &   25 &  36/\ldots &   36 &  0.52 \\
NGC 5338 &               SB$0^{0}$: &  12.8 & $-16.70$ &   95 &   68 &   76 &     26 &  125 &   11 &  15/ 15 &   15 &  0.46 \\
NGC 7280 &            SAB(r)$0^{+}$ &  24.3 & $-19.16$ &   72 &   48 &   66 & \ldots &   55 &    9 &  29/ 27 &   21 &  0.40 \\
NGC 7743 &          (R)SB(s)$0^{+}$ &  20.7 & $-19.49$ &  105 &   28 &   91 &     45 &   95 &   31 &  72/ 58 &   37 &  0.37 \\
  IC 676 &          (R)SB(r)$0^{+}$ &  19.4 & $-18.42$ &   15 &   47 &   74 &     15 &  164 &   13 &  40/ 34 &   18 &  0.72 \\
\multicolumn{13}{c}{Virgo S0 Galaxies} \\
NGC 4267 &            SB(s)$0^{-}$? &  15.3 & $-19.25$ &  127 &   25 &   97 &     28 &   33 &   18 &  28/ 26 &   26 &  0.21 \\
NGC 4340 &             SB(r)$0^{+}$ &  15.3 & $-18.90$ &   95 &   50 &  105 &     53 &   31 &   39 &  51/ 48 &   48 &  0.39 \\
NGC 4371 &             SB(r)$0^{+}$ &  15.3 & $-19.32$ &   92 &   58 &  119 &     37 &  167 &   34 &  42/ 40 &   40 &  0.26 \\
NGC 4477 &           SB(s)$0^{0}$:? &  15.3 & $-19.69$ &   80 &   33 &  114 &     36 &   12 &   25 &  45/ 37 &   37 &  0.35 \\
NGC 4596 &             SB(r)$0^{+}$ &  15.3 & $-19.63$ &  120 &   42 &  119 &     40 &   73 &   52 &  75/ 71 &   57 &  0.51 \\
NGC 4608 &             SB(r)$0^{0}$ &  15.3 & $-19.02$ &  100 &   36 &   97 &     29 &   25 &   44 &  59/ 57 &   49 &  0.48 \\
NGC 4612 &            (R)SAB$0^{0}$ &  15.3 & $-19.01$ &  143 &   44 &   74 & \ldots &   83 &   17 &  24/ 20 &   20 &  0.22 \\
NGC 4754 &            SB(r)$0^{-}$: &  16.8 & $-19.78$ &   23 &   61 &  137 &     36 &  142 &   23 &  27/ 30 &   27 &  0.23 \\
\multicolumn{13}{c}{S0/a Galaxies} \\
NGC 2681 & (R$^{\prime}$)SAB(rs)0/a &  17.2 & $-20.20$ &  140 &   18 &  109 &     27 &   30 &   50 &  75/ 60 &   60 &  0.23 \\
NGC 4245 &                 SB(r)0/a &  12.0 & $-18.28$ &  173 &   38 &   87 &     30 &  137 &   37 &  56/ 42 &   42 &  0.48 \\
NGC 4643 &                SB(rs)0/a &  18.3 & $-19.85$ &   55 &   38 &   93 &     54 &  133 &   50 &  69/ 62 &   62 &  0.45 \\
NGC 4665 &                 SB(s)0/a &  10.9 & $-18.87$ &  120 &   26 &  114 &     37 &    4 &   45 &  99/ 65 &   65 &  0.51 \\
NGC 4691 &              (R)SB(s)0/a &  15.1 & $-19.43$ &   30 &   38 &   85 &     29 &   82 &   30 &  69/ 55 &   45 &  0.64 \\
NGC 5701 &             (R)SB(rs)0/a &  21.3 & $-19.97$ &   45 &   20 &  128 & \ldots &  177 &   40 &  58/ 67 &   58 &  0.37 \\
NGC 5750 &                 SB(r)0/a &  26.6 & $-19.94$ &   65 &   62 &   91 &     22 &  121 &   20 &  24/ 24 &   22 &  0.37 \\
NGC 6654 &   (R$^{\prime}$)SB(s)0/a &  28.3 & $-19.65$ &    0 &   44 &   79 & \ldots &   17 &   26 &  47/ 38 &   38 &  0.51 \\
UGC 11920 &                   SB0/a &  18.0 & $-19.71$ &   50 &   52 &   72 &     38 &   45 &   26 &  39/\ldots &   39 &  0.51 \\
\multicolumn{13}{c}{Sa Galaxies} \\
 NGC 718 &                  SAB(s)a &  22.6 & $-19.43$ &    5 &   30 &   71 &     17 &  152 &   20 &  33/ 30 &   30 &  0.23 \\
NGC 1022 &     (R$^{\prime}$)SB(s)a &  18.1 & $-19.46$ &  174 &   24 &   72 &     24 &  115 &   19 &  33/ 22 &   22 &  0.51 \\
NGC 2273 &                  SB(r)a: &  27.3 & $-20.11$ &   50 &   50 &   97 &     30 &  116 &   14 &  17/ 21 &   17 &  0.43 \\
NGC 3185 &                (R)SB(r)a &  17.5 & $-18.61$ &  140 &   48 &   71 &     20 &  114 &   31 &  34/ 32 &   32 &  0.58 \\
NGC 3729 &                   SB(r)a &  16.8 & $-19.35$ &  170 &   50 &   85 &     24 &   26 &   23 &  27/ 26 &   26 &  0.66 \\
NGC 4045 &                  SAB(r)a &  26.8 & $-19.70$ &   90 &   48 &   81 &     22 &   18 &   18 &  22/ 20 &   20 &  0.30 \\
NGC 4314 &                  SB(rs)a &  12.0 & $-19.12$ &   65 &   25 &  125 &     30 &  146 &   67 &  90/111 &   80 &  0.64 \\
NGC 5377 &                (R)SB(s)a &  27.1 & $-20.29$ &   25 &   59 &  111 & \ldots &   45 &   58 &  78/\ldots &   67 &  0.66 \\
\multicolumn{13}{c}{Sab Galaxies} \\
NGC 3049 &                 SB(rs)ab &  20.2 & $-18.65$ &   26 &   51 &   66 &     15 &   27 &   38 &  60/\ldots &   38 &  0.80 \\
NGC 3368 &                SAB(rs)ab &  10.5 & $-20.37$ &  172 &   50 &  228 & \ldots &  115 &   61 &  80/ 75 &   75 &  0.40 \\
NGC 3504 &              (R)SAB(s)ab &  22.3 & $-20.29$ &  149 &   22 &   81 &     22 &  143 &   29 &  45/ 41 &   34 &  0.60 \\
NGC 4151 & (R$^{\prime}$)SAB(rs)ab: &  15.9 & $-20.70$ &   22 &   20 &  189 &     79 &  130 &   65 & 100/ 90 &   90 &  0.50 \\
NGC 4319 &                  SB(r)ab &  23.5 & $-19.26$ &  135 &   42 &   89 &     12 &  152 &   15 &  22/ 17 &   17 &  0.51 \\
NGC 4725 &                 SAB(r)ab &  12.4 & $-20.69$ &   40 &   42 &  321 &    121 &   50 &  118 & 130/170 &  125 &  0.67 \\
NGC 6012 &              (R)SB(r)ab: &  26.7 & $-19.78$ &   45 &   33 &   63 &     42 &  171 &   30 &  99/ 47 &   36 &  0.55 \\
\hline
    \end{tabular}

\medskip

$R_{25}$ is one-half of the corrected 25th-magnitude diameter $D_{0}$
from RC3, and $h$ is the outer-disc exponential scale length for
galaxies (see Section~\ref{sec:measure-discs}).  The different
measurements of bar size (\amax, \aten, \amin, and \lbar) are
discussed in the text (Section~\ref{sec:measure-bars}); \amax{} and
\lbar{} can be considered lower and upper limits, respectively, for
bar size.  \emax{} is the bar's maximum isophotal ellipticity.  All
disc and bar measurements (except for $R_{25}$) are made from $R$-band
images, except as noted in the Appendix.  Hubble types are from RC3;
$M_{B}$ is based on distance (usually from LEDA; see text for
exceptions) and $B_{tc}$ from LEDA.
   
\end{minipage}
\end{table*}

\setcounter{table}{0}
\begin{table*}
\begin{minipage}{126mm}
    \caption{Continued}
    \begin{tabular}{llrrrcrrrrrrr}
\hline
Name & Type & Distance & $M_{B}$ &  \multicolumn{4}{l}{Outer Disc}
     & \multicolumn{5}{l}{Bar} \\
     &      &          &         &  PA & $i$ & $R_{25}$ & $h$~
     & PA & \amax{} & \amin/\aten{} & \lbar{} & \emax{} \\
     &      & (Mpc)    &          & ($\degr$) & ($\degr$)
     & ($\arcsec$) & ($\arcsec$) & ($\degr$) & ($\arcsec$) 
     & ($\arcsec$)~~~ & ($\arcsec$) & \\
\hline
\multicolumn{13}{c}{Sb Galaxies} \\
NGC 2712 &                  SB(r)b: &  26.5 & $-19.88$ &   10 &   59 &   87 &     20 &   32 &   22 &  27/ 27 &   24 &  0.64 \\
NGC 3351 &                   SB(r)b &  10.0 & $-19.94$ &   13 &   56 &  222 & \ldots &  112 &   52 &  68/ 68 &   58 &  0.42 \\
NGC 3485 &                  SB(r)b: &  20.0 & $-19.03$ &    5 &   26 &   69 &     22 &   45 &   20 &  35/ 34 &   24 &  0.64 \\
NGC 3507 &                   SB(s)b &  14.2 & $-19.21$ &   90 &   27 &  102 &     25 &  112 &   26 &  37/ 33 &   29 &  0.52 \\
NGC 3982 &                 SAB(r)b: &  18.0 & $-19.63$ &   17 &   30 &   70 & \ldots &   10 &    4 &   5/  6 &    5 &  0.25 \\
NGC 4037 &                 SB(rs)b: &  13.5 & $-17.79$ &  150 &   32 &   75 &     34 &   11 &   27 &  38/ 42 &   33 &  0.62 \\
NGC 4102 &                 SAB(s)b? &  14.4 & $-19.22$ &   38 &   55 &   91 & \ldots &   67 &   10 &  15/\ldots &   15 &  0.45 \\
NGC 4699 &                 SAB(rs)b &  18.9 & $-21.37$ &   37 &   42 &  114 &     13 &   50 &   13 &  19/ 16 &   16 &  0.46 \\
NGC 4995 &                  SAB(r)b &  23.6 & $-20.41$ &   93 &   47 &   74 &     17 &   26 &   16 &  24/ 22 &   19 &  0.34 \\
NGC 5740 &                 SAB(rs)b &  22.0 & $-19.67$ &  161 &   60 &   89 &     17 &  123 &   12 &  14/ 22 &   14 &  0.47 \\
NGC 5806 &                  SAB(s)b &  19.2 & $-19.67$ &  166 &   58 &   93 &     30 &  175 &   37 &  95/ 82 &   38 &  0.62 \\
NGC 5832 &                 SB(rs)b? &   9.9 & $-17.15$ &   45 &   55 &  111 &     21 &  159 &   26 &  33/ 34 &   30 &  0.29 \\
NGC 5957 &    (R$^{\prime}$)SAB(r)b &  26.2 & $-19.36$ &  100 &   15 &   85 & \ldots &   97 &   24 &  37/ 31 &   27 &  0.53 \\
NGC 7177 &                  SAB(r)b &  16.8 & $-19.79$ &   83 &   48 &   93 &     17 &   13 &   10 &  14/ 13 &   11 &  0.39 \\
 IC 1067 &                   SB(s)b &  22.2 & $-18.82$ &  110 &   36 &   64 &     15 &  151 &   19 &  23/ 23 &   19 &  0.64 \\
UGC 3685 &                  SB(rs)b &  26.8 & $-19.51$ &  119 &   31 &   99 &     45 &  131 &   25 &  35/ 30 &   27 &  0.59 \\
\hline
    \end{tabular}

\end{minipage}
\end{table*}

\subsection{Measuring the Sizes of Bars} 
\label{sec:measure-bars}

There is no standard way to measure the length of a bar, either for
real galaxies or for simulations.  Methods which have been used for
real galaxies include: visual estimation directly from images
\citep[e.g.,][]{kormendy79,m95}; fitting ellipses to the galaxy
isophotes, with bar length usually determined from a maximum in the
ellipticity
\citep[e.g.,][]{wozniak91,w95,jungwiert97,laine02,sheth03}; various
measurements based on Fourier analysis of the galaxy image, using
either the bar-interbar luminosity contrast
\citep[e.g.,][]{ohta90,aguerri00} or the phase angle
\citep[e.g.,][]{quillen94}; and measurements using the major-axis
profile of the bar \citep[e.g.,][]{seigar98,chapelon99}.  There is
similar variation in how bars are measured even when the galaxy is
readily accessible; i.e., in $n$-body simulations -- compare, for
example, \citet{debattista00}, \citet{athan-m02}, and \citet{vk03}.
As Athanassoula \& Misiriotis demonstrate, different methods applied
to the same (model) galaxies can lead to variations of $\sim15$--35\%
in measured length.\footnote{Based on the mean and standard deviations
from their Table~1.}

After some experimentation, I settled on two measurements, an approach
I also used for the outer and inner bars of double-barred galaxies
\citep{erwin04-db}.  These can be thought of as lower and upper limits
on the bar size.  The lower-limit measurement is \amax, the semimajor
axis of maximum ellipticity in the bar region, which is useful
primarily because it is common and reproducible.  In some cases, there
is no clear ellipticity peak associated with the bar; but a
corresponding extremum in the \textit{position angles} can often be
found which serves the same purpose; examples include NGC~2880 and
NGC~4143 \citep[see][]{erwin-sparke03}.  It is important to stress
that I identify \amax{} with the maximum in ellipticity (or maximum
deviation in position angle) \textit{closest to the end of the bar,
but still inside the bar}.  In some cases, particularly when there are
strong dust lanes and/or star formation, the inner isophotes can
become highly distorted and \textit{more} elliptical than the bar
proper; examples include NGC~2787, IC~676, and NGC~4691 (Erwin \&
Sparke 2003).\footnote{This can happen even in the near-IR:
\citet{ls02} report an unusually small $\amax = 18\arcsec$ for
NGC~4691, from ellipse fits to 2MASS images.} In other cases, the bar
merges so smoothly into spiral arms further out that the ``obvious''
maximum in ellipticity occurs well outside the bar and is due to
spiral arms or a ring.  Examples of this include NGC~3185 and NGC~7743
(Erwin \& Sparke 2003); another good example, albeit a galaxy not in
this study, is NGC~4303 \citep[see the discussion in][]{erwin04-db}.

Despite the relative simplicity and common use of \amax, there is good
reason to believe that it underestimates the true length of the bar.
This has been pointed out by several authors
\citep{w95,erwin-sparke03,lsr02}, and \citet{athan-m02} found that it
generally provided the smallest estimates of bar length in their
$n$-body simulations.  Thus there is a need for a second
(``upper-limit'') measurement, which I refer to as the bar's
``length'' \lbar{}.  This is based on the approach of
\citet{erwin-sparke03}, where the bar length was defined as the
\textit{minimum} of two ellipse-fit measures: the first minimum in
ellipticity outside the bar's peak ellipticity (\amin), or the point
at which the position angles of the fitted ellipses differ by $\geq
10\degr$ from the bar's position angle.  To their definition, I have
added a qualification: if the bar is surrounded by a ring or spiral
arms, and the size of the ring (or arms, where they intersect the bar)
is smaller than either $\amin$ or $\aten$, then I adopt the
ring/spiral size.  This is because chance combinations of orientation
and projection acting on the ring or spirals can lead to ellipse-fit
profiles that place $\amin$ and $\aten$ well \textit{outside} the bar.
Since there is no indication in any of these galaxies that the bar
extends beyond the ring or surrounding spirals, it makes sense to use
the latter as an upper limit on bar size.  Table~\ref{tab:galaxies}
lists \amax, \amin, \aten, and the adopted \lbar{} for each galaxy; if
\lbar{} is smaller than either \amin{} or \aten, then this means that
\lbar{} was derived using rings or spiral arms.

The \lbar{} measurement is perhaps less consistent and accurate than
\amax{} (it is prone to strongly overestimate bar length in face-on
galaxies lacking rings or spiral arms), but may give a better measure
of the bar's ``true'' length -- that is, where the bar distortion
finally gives way to the outer disc or spiral structure.  The only
galaxy where it clearly fails is NGC~4203, which is face-on and
lacking in any spiral arms or rings, so that $\amin = 46\arcsec$ even
though $\amax = 13\arcsec$; consequently, I exclude that galaxy from
statistics using \lbar.

\begin{figure}
\includegraphics[scale=0.45]{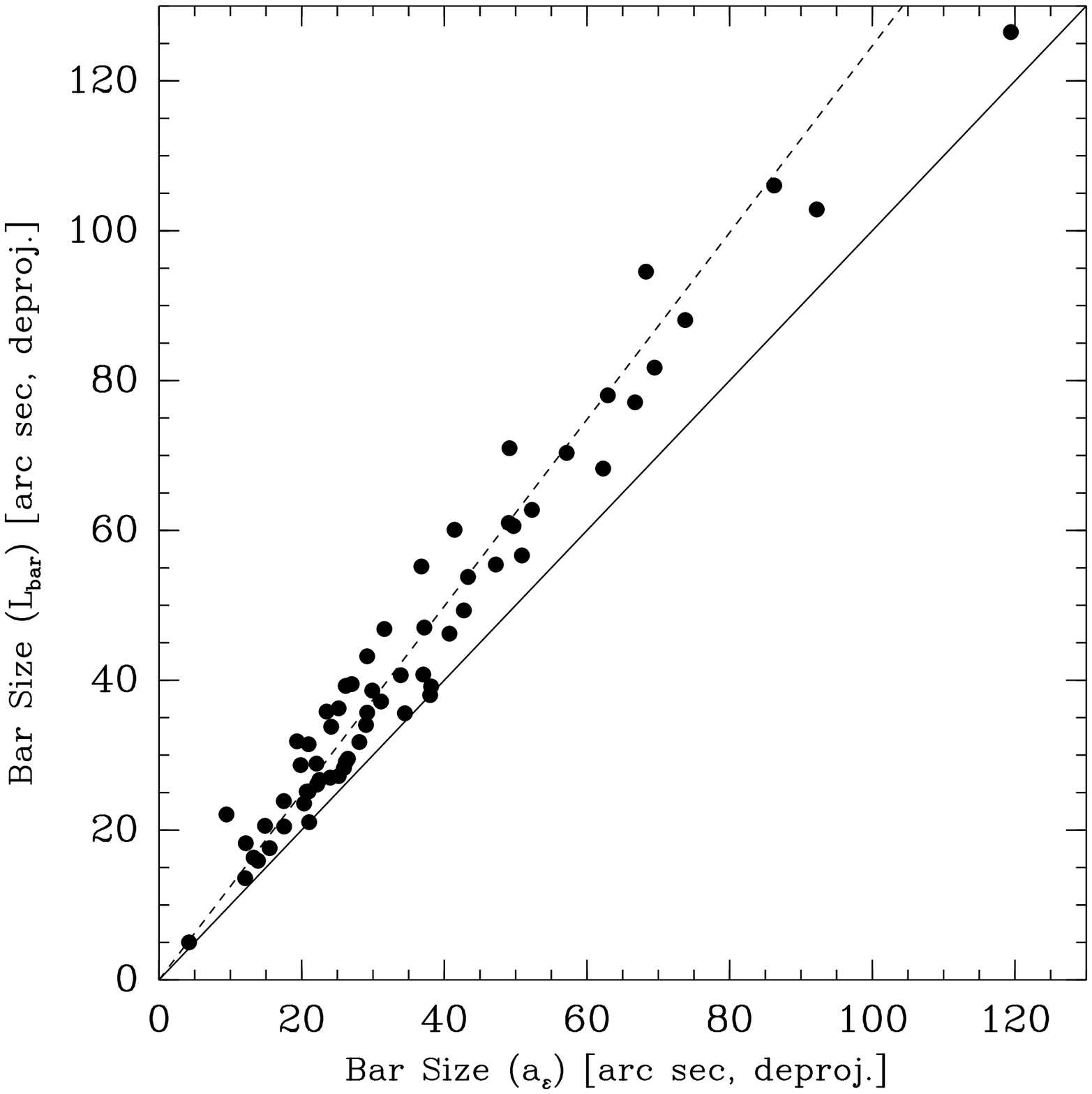}
\caption{Correlation between two deprojected measurements of bar
semimajor axis -- semimajor axis at maximum isophotal ellipticity
($\amax$) and bar length (\lbar{}; see text for definition) -- for
S0--Sb galaxies.  The dashed line indicates the mean ratio of the two 
measurements: $\lbar = 1.25 \amax$.}
\label{fig:amax-v-lbar}
\end{figure}

In practice, the two measurements are extremely well correlated
(Figure~\ref{fig:amax-v-lbar}); the Pearson and Spearman correlation
coefficients are $r = 0.96$ and $r_{s} = 0.95$,
respectively.\footnote{As a reminder, the Pearson coefficient measures
the strength of \textit{linear} correlations; the Spearman coefficient
measures general correlations and is usually considered more robust
against outliers; see, e.g., \citet{nr}.} The mean (deprojected) ratio
of $\amax / \lbar$ is 0.80.  This is not too far from the mean ratio
(0.73) of sizes for the $L_{b/a}$ and \lph{} measurements of
\citet{athan-m02}, which suggests that those two $n$-body bar
measurements are a good match to \amax{} and \lbar{} (see
Section~\ref{sec:sims}).

The position angles of the bars are also needed, since I use them to
deproject bar sizes.  As shown by \citet{erwin-sparke03}, ellipse fits
are a problematic source for bar position angles.  Their Table~5 lists
11 large-scale bars whose position angles differ from those given by
the ellipse fits by more than 5\degr; see their Figures~5 and 6 for
examples.  Thus, bar position angles are \textit{always} checked
against the images, and the position angle determined from the
isophotes and unsharp masking is preferred to the ellipse-fit position
angle if the two differ by more than a couple of degrees.

Finally, the inclination and line-of-nodes of the galaxy discs need to
be determined.  The easiest approach is to use the RC3 axis
ratios and position angles and assume that the outer disc is circular.
Unfortunately, this is by no means the most accurate way, especially
for early-type barred galaxies.  This is because the RC3 axis ratios
sometimes reflect bar-related features such as inner rings, lenses,
and outer rings, which are more common in early-type disc galaxies and
which are \textit{not} always intrinsically circular
\citep{buta86,buta95}.  So I determined the outer disc orientation,
where possible, using kinematic information (e.g., \hi{} maps) and/or
isophotes at diameters larger than $D_{25}$.  For the WIYN Sample
galaxies, I use the values from \citet{erwin-sparke03}, which were
determined using this approach (certain exceptions based on more
recent data are mentioned in the Appendix).  Details for the Virgo S0
and the Sab--Sb galaxies are given in the Appendix.

\subsection{Measuring the Shapes of Bars} 
\label{sec:measure-shapes}

Another way to define a bar is by its ``strength.''  This too lacks
an obvious, universally agreed-upon definition.  The simplest way to
measure a bar's strength is to measure its \textit{shape}, usually
reduced to measuring its ``ellipticity.''  For theorists, this often
means the ellipticity of the bar itself (e.g., that of a Ferrers
ellipsoid), but for observers -- lacking the ability to unambiguously
isolate the bar from other galactic components -- it usually means
measuring the semi-minor axis of the \textit{isophote} defined by the
bar length (usually \amax) and comparing it with the semimajor axis.
This is approximately the method used by \citet{m95} to define bar
strengths, and also by \citet{shlosman00} and \citet{laine02} for
fitted ellipses defining bars.  For comparison, if no other reason, it
makes sense to do the same.

A more complex approach, which attempts to estimate the
non-axisymmetric gravitational influence of the bar, is that of
\citet{buta01}.  Unfortunately, this generally requires near-IR
images, and assumes that the entire galaxy is flat with a constant
scale height.  For bars in late-type galaxies, where the bulge is
small or even absent, this is probably reasonable; but for early-type
galaxies, large bulges -- and possibly multiple disc components with
different thicknesses -- make this a questionable assumption.  (More
recently, \nocite{lsbv04}Laurikainen et al.\ 2004 have an included a
spherical bulge component in the modeling process, which alleviates
some of the problems.)  Happily, \citet{lsr02} find that the bar
strength measured this way correlates quite well with bar ellipticity.

\subsection{Measuring the Sizes of Discs} 
\label{sec:measure-discs}

Galaxies come in many sizes, and what might be a large bar in one
galaxy would be small in another.  Thus, although absolute
measurements of bars size (in kpc) are useful, we also need some kind
of relative measurement.  What should we compare bar sizes with?

The simplest approach is to follow \citet{ee85} and \citet{m95}:
compare the bar size to the optical disc size $D_{25}$, which is
available for all the galaxies.  Since I measure bar semimajor axes,
I use $R_{25} = D_{25}/2$ for the disc size.  To be consistent with
Martin's measurements, I use the $D_{0}$ values from RC3, which are
corrected for inclination (usually a very small effect) and for
Galactic extinction.

Another useful measurement is the exponential scale length of the
disc.  \citet{combes93} argued that bars in late-type galaxies should
extend to approximately one disc scale length, and \citet{laine02}
suggested that the correlation they observed between bar size
($\amax$) and $D_{25}$ implied that bars ``extend to a fixed number of
radial scale lengths in the disk.''  Bar sizes in terms of disc scale
lengths are also much easier to compare with simulations, since
exponential scale lengths for $n$-body discs are easily measured.

For those galaxies in which an outer exponential disc can be
identified, I derive its slope by fitting the region \textit{outside
the bar}, using an azimuthally averaged surface-brightness
profile.\footnote{That is, a profile obtained with concentric, similar
ellipses using the ``Outer Disc'' position angle and ellipticity from
Table~\ref{tab:galaxies}.} This is the classic approach of ``marking
the disc'' by eye.  In principle, more accurate scale lengths might be
derived by performing a bulge-disc decomposition, so that the
contribution of bulge light to the outer disc profile is accounted
for.  I do \textit{not} attempt this, however, since many of these
galaxies are strongly barred and/or contain luminous central
structures apart from the bulge (secondary bars, inner discs, or
nuclear rings).  In extreme cases, these non-bulge components can
dominate the interior light \citep{erwin03-id}, and attempting to fit
them with, e.g., a de Vaucouleurs or S\'ersic profile could assign too
much light at large radii to the ``bulge'' and distort the disc fit.
Since I am deliberately fitting only the region outside the bar, the
effect of bulge light in most cases is minimized.  In any event,
\citet{dejong96} found that the change in derived scale length between
marking the disc and more sophisticated decomposition techniques was
typically only a few percent.  A comparison of my disc scale lengths
with those of \citet{bba98}, which were obtained via bulge-disc
decompositions, show a similarly small variation; see
Section~\ref{sec:m95} below.  Details for unusual cases are provided
in the Appendix, along with notes for galaxies where severely
non-exponential outer discs prevented determining an exponential scale
length.

\begin{figure}
\includegraphics[scale=0.5]{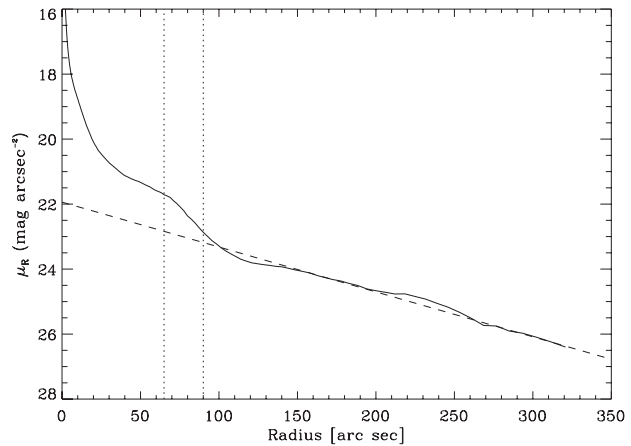}
\caption{Azimuthally averaged $R$-band profile for NGC~4151.  The
vertical, short-dashed lines indicate deprojected bar-size
measurements \amax{} and \lbar; the diagonal, long-dashed line is an
exponential fit to the disc region outside the bar ($r > 145\arcsec$, 
excluding the ring excess at $r \sim 200$--260\arcsec).
Note the excess light at $r \sim 40$--100\arcsec, which is primarily
due to the bar.}
\label{fig:n4151}
\end{figure}

Another reason for fitting the outer disc only is shown in
Figure~\ref{fig:n4151}.  Although it is sometimes argued that bars
vanish from the surface-brightness profile when it is averaged,
leaving behind only the underlying exponential disc
\citep[e.g.][]{ee85,ohta90}, this is not always true: the bar can
sometimes appear as a clear, significant excess above the (outer)
exponential profile \citep[see also][]{dejong96}.  None the less, if
we exclude the bar region and any excess just outside it, we can still
distinguish an unambiguous outer exponential profile in such galaxies
(see Erwin, Pohlen, \& Beckman, in preparation, for details).

\begin{figure}
\includegraphics[scale=0.5]{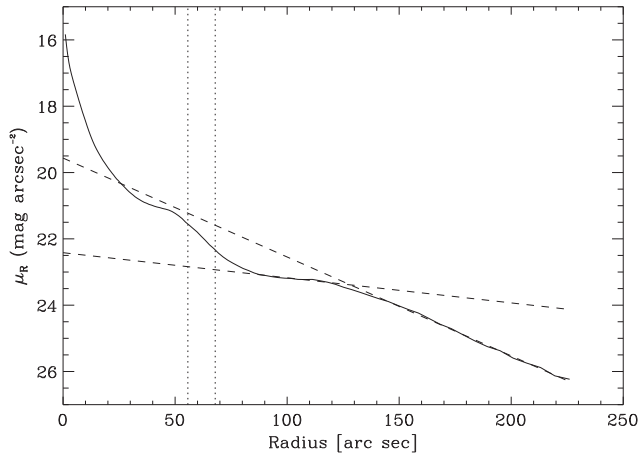}
\caption{As for Figure~\ref{fig:n4151}, but showing the profile of 
NGC~3945, which has \textit{two} exponential zones outside the bar 
(an example of a Freeman Type~II profile).  Because there is no 
single, well-defined exponential zone in such galaxies, I do not 
compute bar sizes relative to disc scale lengths for them.}
\label{fig:n3945}
\end{figure}

However, in other galaxies the outer disc does not have a single
exponential profile.  For 16 galaxies, the profile outside the bar is
what \citet{freeman70} termed a ``Type~II'' profile: it is divided
into \textit{two} (usually) exponential zones: a shallow inner zone
and a steeper outer zone (see Figure~\ref{fig:n3945} for an example).
In such cases, it is not at all clear which of the two zones -- if
either -- should be considered the ``true'' outer disc.  In at least
some cases (e.g., NGC~2859, 3412, 2962, 5701, and 6654), the inner
zone is extremely narrow and/or shallow in slope, or even
\textit{increasing} in brightness with radius.  The simplest solution
is to ignore these more extreme ``outer'' Type~II profiles.  (Five
other galaxies have Type~II profiles with a deficit \textit{inside}
the bar, so there is still a single exponential profile
\textit{outside} the bar; these are similar to profiles produced in
some $n$-body simulations -- e.g., \nocite{vk03}Valenzuela \& Klypin
2003).

There are additional galaxies where the surface brightness profile at
large radii is shallower than that of disc immediately outside the
bar; these are the ``Type III'' or ``anti-truncation'' profiles
discussed in \citet{erwin05-type3}.  Because these galaxies
\textit{do} have an extended, well-defined exponential zone outside
the bar, I include their scale length measurements.  The presence of
extended light at large radius may contribute to a subtle selection
effect, which I discuss in Section~\ref{sec:early-sizes}.

\subsection{Comparing Measurements: Early- and Late-Type Galaxies} 
\label{sec:m95}

The bar sizes of the S0--Sb galaxies are measured using a combination
of ellipse fits and direct inspection of $R$-band images, supplemented
by near-IR images when dust is strong.  The bar sizes of \citet{m95},
on the other hand, are based on measurements made on blue photographic
prints.  How consistent are these measurements?  And which of my two
measurements (\amax{} and \lbar) is a better match to Martin's single
bar-size measurement?

\begin{figure}  
\includegraphics[scale=0.45]{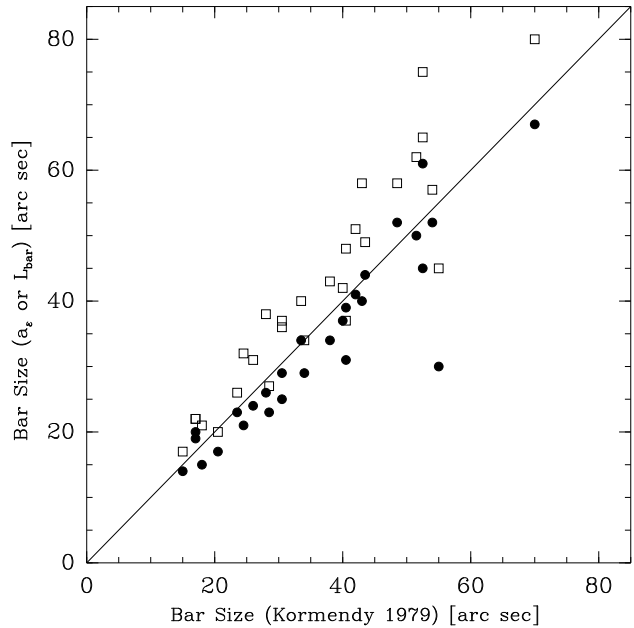}
\caption{Comparison of observed (i.e., projected) bar size
measurements between this study and \citet{kormendy79}: \amax{} values
are shown with filled circles, \lbar{} with open boxes.}
\label{fig:k79bars}
\end{figure}

Unfortunately, there are almost no galaxies in common between the two
samples, even among the Sb subset (the three shared galaxies are
NGC~3351, 3485, and 4725, where Martin finds $a = 46\arcsec$,
18\arcsec, and 109\arcsec, respectively; these values are only
slightly smaller than my $\amax = 52\arcsec$, 23\arcsec, and
118\arcsec).  This makes it difficult to tell which of the two,
\amax{} and \lbar{}, is a better match to Martin's visual estimates.
Fortunately, Martin compared his measurements with the visual bar-size
measurements of \citet{kormendy79} -- and there \textit{are} numerous
overlaps between Kormendy's sample and mine.\footnote{There is not
enough information about bar orientation in Kormendy's Table~1 to
allow reliable deprojections of his bar sizes; in addition, his sample
is limited to SB galaxies.} In Figure~\ref{fig:k79bars}, I compare
\amax{} and \lbar{} measurements with those of Kormendy for galaxies
in common.  Kormendy's measurements generally sit in between \amax{}
and \lbar{}; in fact, the best agreement is with the \textit{mean} of
$\amax$ and \lbar{}, which I will refer to as \lavg.  Since Martin
found excellent agreement between Kormendy's measurements and
\textit{his} for galaxies in common between \textit{their} samples,
this suggests that \lavg{} is a reasonable match to Martin's bar
lengths.  (The results discussed below do not change significantly if
I compare Martin's measurements with \amax{} instead.)

The disc scale lengths for the galaxies from \citet{m95}, when
available, are taken from \citet[][hereafter BBA98]{bba98}.  This is
the only study with large numbers of scale lengths which overlaps
significantly with both the early-type galaxies in my sample
\textit{and} the later-type galaxies in Martin's sample.  The BBA98
fits differ from mine in using major-axis cuts from $V$-band
photographic images, and in using bulge-disc decompositions with an
optional ``hole'' in the disc.  Since the ``bulge'' component (i.e.,
the central excess over the exponential disc, which BBA98 model with a
de Vaucouleurs profile) in late-type galaxies is generally smaller and
less luminous than in earlier types, the effect on the disc fit is
reduced.

The optional hole in their disc model, which is intended to account
for Type~II profiles, introduces a new problem, however: it forces the
bulge component to account for all the inner light (including, e.g.,
the inner part of the disc and the bar).  Since the bulge model is not
truncated, it ends up contributing more light at large radii than the
true bulge probably does, and can thus distort the disc fit.  The
``disc with hole'' fits also obviously indicate possible Type~II
profiles, which, as noted in Section~\ref{sec:measure-discs}, are
problematic for estimating scale lengths.  However, because they use
major-axis cuts rather than azimuthally averaged profiles, at least
some of their Type~II profiles turn out to be Type~I when azimuthally
averaged.  This generally happens when the bar is oriented at an
intermediate angle or perpendicular to the major-axis cut, and
probably signals the transition between the outer disc and, e.g, a
lens in which the bar is embedded (e.g., NGC~2787; see Erwin, Pohlen,
\& Beckman, in preparation, and also Ohta et al.\ 1990).

\begin{figure}  
\includegraphics[scale=0.45]{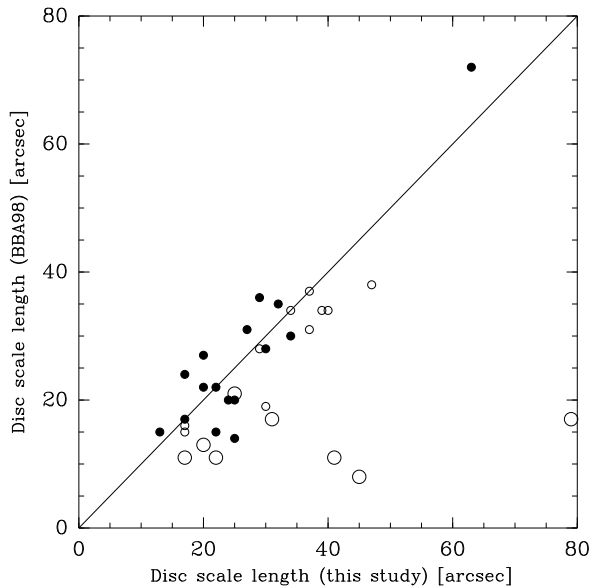}
\caption{Comparison of exponential disc scale lengths for galaxies in
common between this study and BBA98 \citep{bba98}.  Different symbols
indicate different types of fits from Baggett et al.: filled circles
are standard discs (no holes), small hollow circles are fits using
discs with holes having radii $< 2$ times disc scale length $h$, and
large hollow circles are fits with large holes ($r_{\mathrm{hole}} > 2
h$).  The agreement is generally good except when the BBA98 fits have
large holes.}
\label{fig:bba98h}
\end{figure}

To evaluate how well the BBA98 disc scale lengths compare with mine,
Figure~\ref{fig:bba98h} plots scale lengths for galaxies in common
between this study and BBA98.  There is a fair amount of scatter, but
the agreement is generally good \textit{if} I reject BBA98 scale
lengths when the hole radius is more than twice the scale length.
This criterion would also reject 6 of the 9 Type~II profiles from my
sample which are also in BBA98 (not plotted), so I use the BBA98 scale
lengths only when their fit has no hole or a hole with radius $< 2 h$.

The comparison in Figure~\ref{fig:bba98h} is also useful as a test of
how well scale lengths measured directly from the profile (``marking
the disc'') compare with those derived from a bulge-disc
decomposition, as done by BBA98.  Except for cases where BBA98 used
very large holes in their discs (large open symbols), the agreement is
rather good.  There is a very weak systematic trend: my scale lengths
are on average 4\% larger than the BBA98 scale lengths.  But since the
\textit{relative} scatter (absolute value) is 17\%, this is not a very
significant difference.

\begin{figure}  
\includegraphics[scale=0.5]{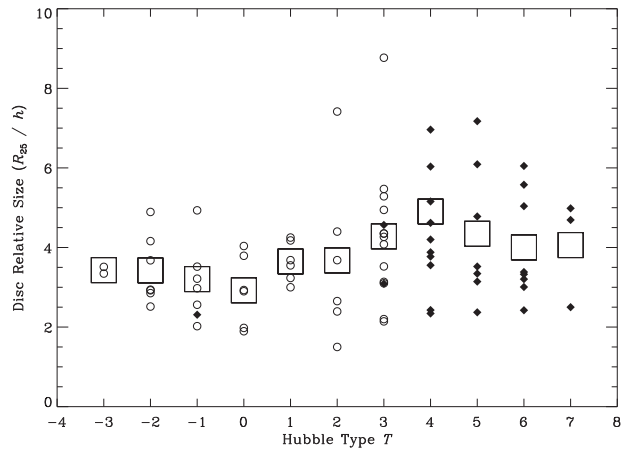}
\caption{Isophotal disc size ($R_{25}$) in terms of the exponential
scale length $h$, as a function of Hubble type.  Galaxies from my
sample are shown with open circles, while galaxies from \citet{m95}
meeting the same selection criteria are shown with filled diamonds;
mean values for each Hubble type are indicated by the large boxes.}
\label{fig:r25h}
\end{figure}

Figure~\ref{fig:r25h} shows the run of $R_{25}/h$ versus Hubble type
for the combined samples.  In general, the early-type galaxies have
$R_{25} \sim 3 h$, while Sb and later-type galaxies have $R_{25} \sim
4 h$, presumably because their discs have more younger stars, and are
thus bluer and brighter.  As I will show in
Section~\ref{sec:early-sizes}, this has some implications for relative
bar sizes.

\section{Bars Sizes, Strengths, and Hubble Type}

In this section I look at how bar sizes, in absolute and relative
terms, vary with Hubble type and with bar strength.  I also
investigate whether and to what degree the size of bars correlates
with galaxy size and luminosity, and with bar strength.  I begin by
discussing the S0--Sb bars from my sample in isolation, both because
my sample is more consistent and complete than that of \citet{m95} and
because I measured \textit{two} bar sizes (\amax{} and \lbar) versus
the single measurement of Martin.  I then add in the later types of
Martin's sample and look at the run of bar sizes along the Hubble
sequence from S0--Sd, and finally examine how bar size relates to bar
strength.

\subsection{Bars in Early-Type Galaxies}
\label{sec:early-sizes}

\begin{figure}  
\includegraphics[scale=0.5]{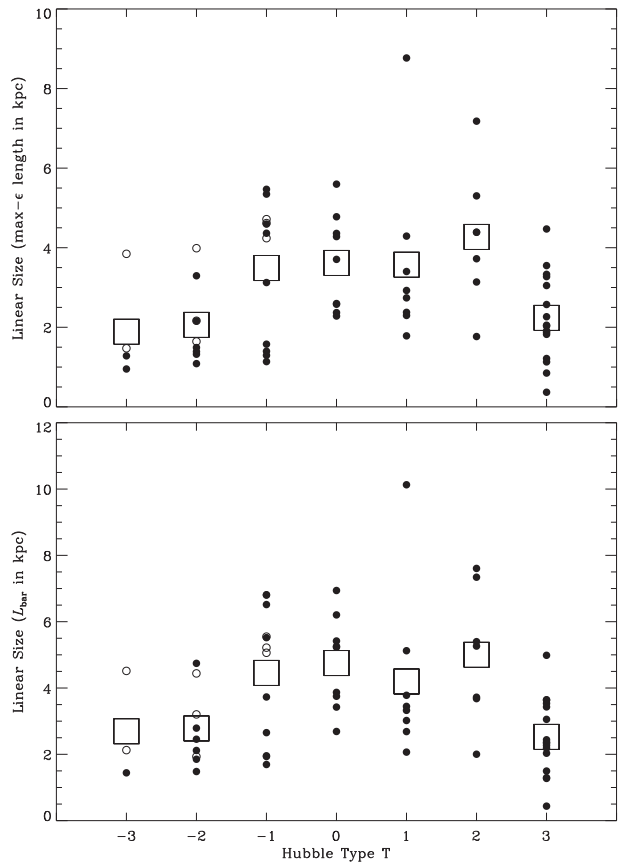}
\caption{Deprojected sizes of S0--Sb bars in absolute terms
(semimajor axis in kpc), using maximum-ellipticity length \amax{}
(top) and \lbar{} (bottom).  Virgo Cluster S0 galaxies are indicated
by hollow circles, with filled circles for field galaxies.  The mean
values for each Hubble type are indicated by the large boxes.}
\label{fig:kpc-size}
\end{figure}

\begin{figure}  
\includegraphics[scale=0.5]{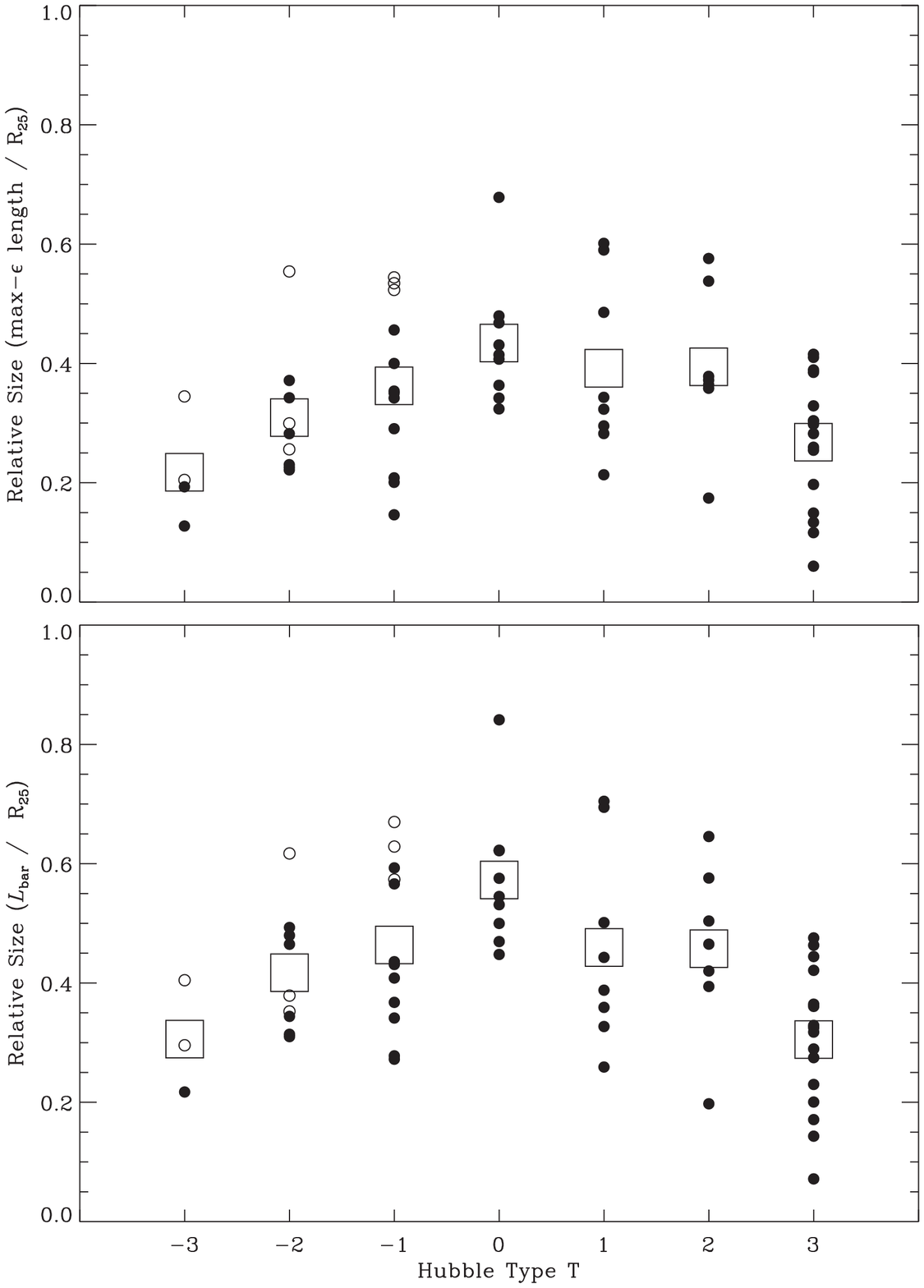}
\caption{As for Figure~\ref{fig:kpc-size}, but now showing sizes of
bars relative to disc radius $R_{25}$.}
\label{fig:r25-size}
\end{figure}

\begin{figure}  
\includegraphics[scale=0.5]{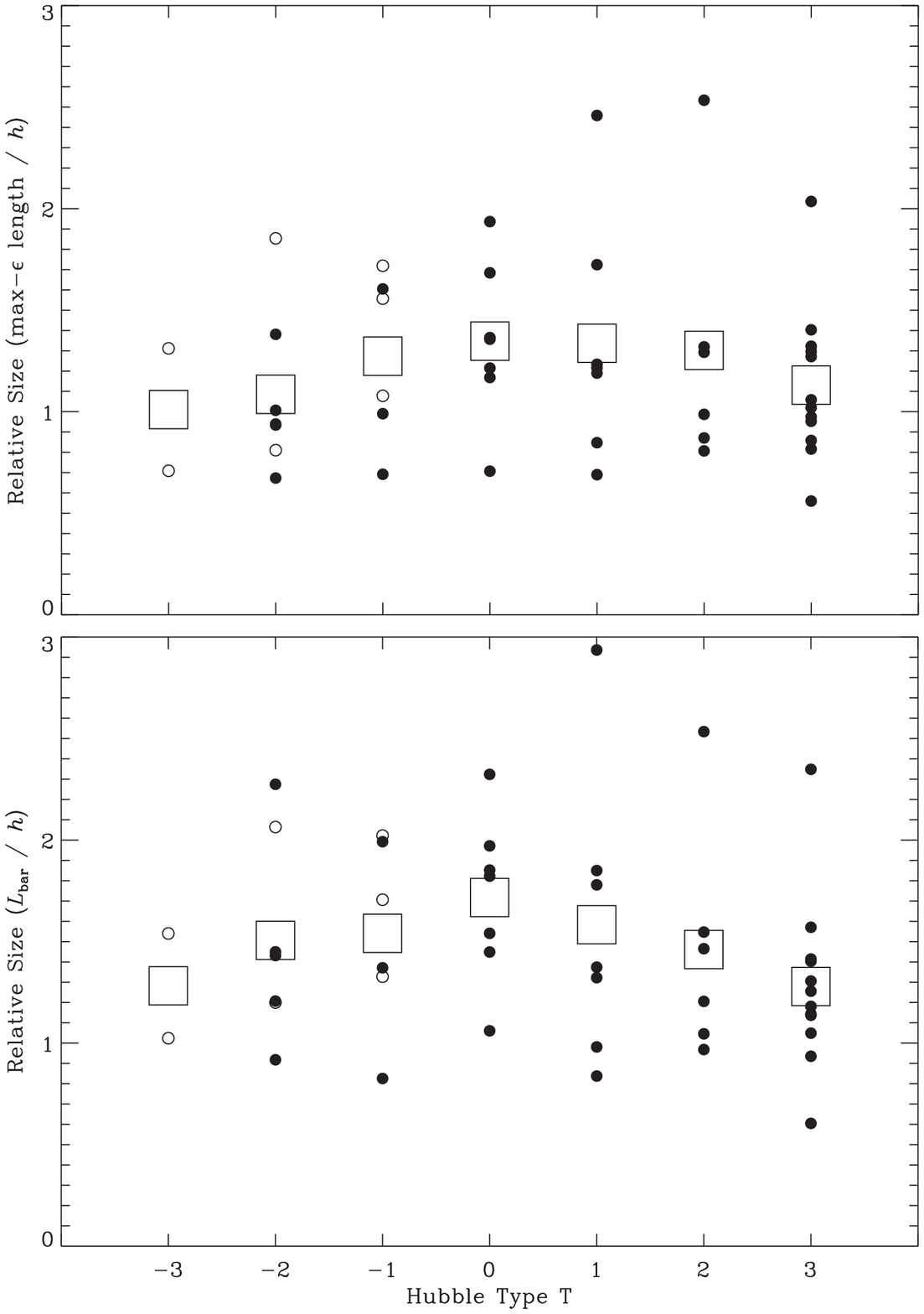}
\caption{As for Figure~\ref{fig:kpc-size}, but now showing sizes of bars
relative to the outer-disc exponential scale length $h$.}
\label{fig:h-size}
\end{figure}

Figure~\ref{fig:kpc-size} shows absolute bars sizes as a function of
Hubble type for the S0--Sb galaxies; Figures~\ref{fig:r25-size} and
\ref{fig:h-size} show the run of bar sizes relative to $R_{25}$ and
the disc scale length $h$.  Table~\ref{tab:sizes} shows mean bar sizes
for different galaxy types, and Tables~\ref{tab:correlations} and
\ref{tab:correlations-h} show the strength of different correlations
between bar sizes and galaxy properties.

\begin{table}
\caption{Mean Bar Sizes for S0--Sb Galaxies}
\label{tab:sizes}
\begin{tabular}{@{}lll}
\hline
Types & Mean \amax{} & Mean \lbar{} \\
\hline
\multicolumn{3}{c}{Absolute Size (kpc)} \\
\hline
All S0--Sb    & $3.0 \pm 1.6$    & $3.7 \pm 1.9$ \\
All S0        & $2.7 \pm 1.6$    & $3.6 \pm 1.8$ \\
S0/a          & $3.6 \pm 1.2$    & $4.8 \pm 1.4$ \\
Sa            & $3.6 \pm 2.2$    & $4.2 \pm 2.6$ \\
Sab           & $4.3 \pm 1.7$    & $5.0 \pm 2.0$ \\
Sb            & $2.2 \pm 1.1$    & $2.5 \pm 1.1$ \\
SB            & $3.2 \pm 1.6$    & $3.8 \pm 1.8$ \\
SAB           & $2.7 \pm 1.7$    & $3.5 \pm 2.0$ \\
\hline
\multicolumn{3}{c}{Fraction of Disc Size $R_{25}$} \\
\hline
All S0--Sb    & $0.34 \pm 0.13$    & $0.42 \pm 0.15$ \\
All S0        & $0.32 \pm 0.13$    & $0.43 \pm 0.13$ \\
S0/a          & $0.43 \pm 0.11$    & $0.57 \pm 0.12$ \\
Sa            & $0.39 \pm 0.15$    & $0.46 \pm 0.16$ \\
Sab           & $0.39 \pm 0.13$    & $0.46 \pm 0.14$ \\
Sb            & $0.27 \pm 0.11$    & $0.31 \pm 0.12$ \\
SB            & $0.37 \pm 0.13$    & $0.45 \pm 0.15$ \\
SAB           & $0.27 \pm 0.11$    & $0.37 \pm 0.14$ \\
\hline
\multicolumn{3}{c}{Fraction of Disc Scale Length $h$} \\
\hline
All S0--Sb    & $1.27 \pm 0.52$    & $1.51 \pm 0.56$ \\
All S0        & $1.16 \pm 0.43$    & $1.44 \pm 0.42$ \\
S0/a          & $1.37 \pm 0.42$    & $1.70 \pm 0.44$ \\
Sa            & $1.43 \pm 0.82$    & $1.63 \pm 0.96$ \\
Sab           & $1.30 \pm 0.72$    & $1.46 \pm 0.64$ \\
Sb            & $1.20 \pm 0.49$    & $1.38 \pm 0.58$ \\
SB            & $1.27 \pm 0.55$    & $1.50 \pm 0.58$ \\
SAB           & $1.29 \pm 0.41$    & $1.54 \pm 0.49$ \\
\hline
\end{tabular}

\end{table}

Bar and disc sizes are well correlated for these galaxies; the
correlations between bar sizes and outer-disc scale length are equally
good.  In fact, when only those galaxies with measured scale lengths
are considered (Table~\ref{tab:correlations-h}), the
\textit{strongest} correlation is with the disc scale length (this may
be biased by the Sb galaxies; see below).  There is also a correlation
between bar size and blue luminosity, as first noted by
\citet{kormendy79}; however, it is clearly not as strong as the
correlations with $h$ and $R_{25}$.  This is reasonable, since bars
are disc phenomena and $M_{B}$ can include variable contributions from
the bulge; in addition, it may be more affected by variations in star
formation and dust extinction.

The correlation between bar size and scale length is strongest for Sb
galaxies (Table~\ref{tab:correlations-h}), but the correlations with
$R_{25}$ and $M_{B}$ are noticeably weaker.  The Sb galaxies are also
odd in having bars which are smaller in absolute and $R_{25}$-relative
size (Figures~\ref{fig:kpc-size} and \ref{fig:r25-size}) -- and yet
almost the \textit{same} size relative to the disc scale length
(Figure~\ref{fig:h-size}).

This curious situation may partly be due to higher levels of recent
star formation in the Sb and later galaxies, which makes their discs
brighter and thus larger -- in blue isophotal size -- for a given
scale length (see Figure~\ref{fig:r25h}).  For the Sb galaxies in my
sample, $R_{25}/h = 3.9 \pm 1.1$ ($3.9 \pm 1.0$ if Sb galaxies from
\nocite{m95}Martin [1995] are included), versus $3.2 \pm 0.9$ for the
S0--Sab galaxies.  For the combined Sb--Sd galaxies, the mean size is
$R_{25}/h = 4.3 \pm 1.7$.  Kolmogorov-Smirnov (K-S) tests suggest that
there is indeed a significant difference in relative disc size between
early and late types, starting with the Sb galaxies: the Sb--Sd
$R_{25}/h$ ratios are inconsistent with those of S0--Sab galaxies ($P
= 0.012$).

Thus, my selection criterion of $D_{25} \geq 2.0\arcmin$ means that
the Sb subsample includes galaxies with smaller $h$ -- and thus bars
with smaller absolute sizes -- than the earlier Hubble types: mean $h
= 2.5 \pm 1.7$ kpc for Sb galaxies versus $3.0 \pm 1.6$ kpc for
S0--Sab galaxies.  If the star formation in Sb and later galaxies is
also more \textit{variable} than in earlier types, then the weaker
correlations of Sb bar size with $R_{25}$ and $M_{B}$ make sense as
well.

An additional possible effect is the existence of galaxies with excess
light at larger radii, relative to the outward projection of the
exponential disc profile.  These are the Type III profiles of
\citet{erwin05-type3}, and they are, in this sample at least,
especially common in both Sb galaxies and S0 galaxies (31\% and 32\%,
respectively, of those Hubble types, compared with 13\% of the
S0/a--Sab galaxies).  Again, if we assume that bar size scales most
strongly with inner-disc $h$, then a diameter-limited selection will
preferentially include S0 and Sb galaxies with smaller disc scale
lengths and smaller bars, when compared with S0/a--Sab galaxies in the
same sample.

Table~\ref{tab:sizes} \textit{does} show a slight tendency for both S0
and Sb bars to be smaller in size relative to disc scale length, but
this is not statistically significant: for example, a K-S test gives
$P = 76$--87\% that Sb and S0/a--Sab bar sizes relative to $h$ come
from the same parent population.  So despite what
Figures~\ref{fig:kpc-size} and \ref{fig:r25-size} seem to suggest, it
is not clear that S0 and Sb bars are really smaller than the bars in
S0/a--Sab galaxies.  If size relative to disc scale length is the most
reliable measuring stick, then there is no significant difference in
bar size over the range S0--Sb.

Although most of the galaxies in my sample are from the field, I did
include eight S0 galaxies from the Virgo Cluster.  Do Virgo S0's have
different bar properties from field S0's?  The Virgo lenticulars
\textit{do} tend to have slightly larger bars than the field S0's
(e.g., mean $\amax/R_{25} = 0.41 \pm 0.15$ versus $0.28 \pm 0.09$;
mean $\amax/h = 1.25 \pm 0.49$ versus $1.06 \pm 0.39$).  However, none
of these differences is statistically significant: K-S tests give
probabilities of 9--88\% for the field and Virgo S0 bar sizes being
drawn from the same parent population.

\begin{table}
\caption{Correlations for S0--Sb Galaxies}
\label{tab:correlations}
\begin{tabular}{@{}lll}
\hline
Correlation & Pearson $r$ ($P$) & Spearman $r_{s}$ ($P$) \\
\hline
\multicolumn{3}{c}{All S0--Sb Galaxies} \\
\hline
\amax{} vs.\ $R_{25}$ & 0.73 ($9.4 \times 10^{-12}$) & 0.69 ($3.9 \times 10^{-10}$) \\
\lbar{} vs.\ $R_{25}$ & 0.72 ($1.4 \times 10^{-11}$) & 0.68 ($6.0 \times 10^{-10}$) \\
\\
\amax{} vs.\ $M_{B}$ & $-0.46$ ($1.3 \times 10^{-4}$) & $-0.51$ ($1.5 \times 10^{-5}$) \\
\lbar{} vs.\ $M_{B}$ & $-0.48$ ($6.4 \times 10^{-5}$) & $-0.53$ ($7.8 \times 10^{-6}$) \\
\hline
\multicolumn{3}{c}{S0 Galaxies} \\
\hline
\amax{} vs.\ $R_{25}$ & 0.81 ($1.8 \times 10^{-6}$) & 0.75 ($2.3 \times 10^{-5}$) \\
\lbar{} vs.\ $R_{25}$ & 0.84 ($2.2 \times 10^{-7}$) & 0.84 ($3.0 \times 10^{-7}$) \\
\\
\amax{} vs.\ $M_{B}$ & $-0.58$ (0.0029) & $-0.59$ (0.0025) \\
\lbar{} vs.\ $M_{B}$ & $-0.62$ (0.0013) & $-0.64$ ($8.1 \times 10^{-4}$) \\
\hline
\multicolumn{3}{c}{S0/a--Sab Galaxies} \\
\hline
\amax{} vs.\ $R_{25}$ & 0.70 ($1.2 \times 10^{-4}$) & 0.64 ($7.2 \times 10^{-4}$) \\
\lbar{} vs.\ $R_{25}$ & 0.69 ($1.7 \times 10^{-4}$) & 0.65 ($5.8 \times 10^{-4}$) \\
\\
\amax{} vs.\ $M_{B}$ & $-0.59$ (0.0026) & $-0.64$ ($7.7 \times 10^{-4}$) \\
\lbar{} vs.\ $M_{B}$ & $-0.64$ ($8.5 \times 10^{-4}$) & $-0.66$ ($4.5 \times 10^{-4}$) \\
\hline
\multicolumn{3}{c}{Sb Galaxies} \\
\hline
\amax{} vs.\ $R_{25}$ & 0.61 (0.013) & 0.58 (0.019) \\
\lbar{} vs.\ $R_{25}$ & 0.60 (0.015) & 0.56 (0.025) \\
\\
\amax{} vs.\ $M_{B}$ & $-0.05$ (0.84) & $-0.15$ (0.59) \\
\lbar{} vs.\ $M_{B}$ & $-0.05$ (0.85) & $-0.15$ (0.58) \\
\hline
\end{tabular}

 \medskip
 Correlations between bar size (\amax, \lbar) and galaxy size
 ($R_{25}$) and blue luminosity ($M_{B}$).  For each correlation
 coefficient, the probability of the correlation being purely due to
 chance is given in parentheses.
\end{table}

\begin{table}
\caption{Correlations for S0--Sb Galaxies with Measured Scale Lengths}
\label{tab:correlations-h}
\begin{tabular}{@{}lll}
\hline
Correlation & Pearson $r$ ($P$) & Spearman $r_{s}$ ($P$) \\
\hline
\multicolumn{3}{c}{All S0--Sb Galaxies} \\
\hline
\amax{} vs.\ $h$ & 0.73 ($6.4 \times 10^{-9}$) & 0.72 ($10.0 \times 10^{-9}$) \\
\lbar{} vs.\ $h$ & 0.75 ($1.8 \times 10^{-9}$) & 0.75 ($1.8 \times 10^{-9}$) \\
\\
\amax{} vs.\ $R_{25}$ & 0.63 ($2.5 \times 10^{-6}$) & 0.53 ($1.2 \times 10^{-4}$) \\
\lbar{} vs.\ $R_{25}$ & 0.59 ($1.5 \times 10^{-5}$) & 0.48 ($5.8 \times 10^{-4}$) \\
\\
\amax{} vs.\ $M_{B}$ & $-0.40$ (0.006) & $-0.39$ (0.0071) \\
\lbar{} vs.\ $M_{B}$ & $-0.42$ (0.0032) & $-0.41$ (0.0044) \\
\hline
\multicolumn{3}{c}{S0 Galaxies} \\
\hline
\amax{} vs.\ $h$ & 0.69 (0.0044) & 0.71 (0.0028) \\
\lbar{} vs.\ $h$ & 0.72 (0.0024) & 0.80 ($3.4 \times 10^{-4}$) \\
\\
\amax{} vs.\ $R_{25}$ & 0.70 (0.004) & 0.69 (0.0048) \\
\lbar{} vs.\ $R_{25}$ & 0.74 (0.0016) & 0.76 ($9.1 \times 10^{-4}$) \\
\\
\amax{} vs.\ $M_{B}$ & $-0.47$ (0.074) & $-0.43$ (0.11) \\
\lbar{} vs.\ $M_{B}$ & $-0.55$ (0.032) & $-0.54$ (0.038) \\
\hline
\multicolumn{3}{c}{S0/a--Sab Galaxies} \\
\hline
\amax{} vs.\ $h$ & 0.76 ($8.7 \times 10^{-5}$) & 0.57 (0.0081) \\
\lbar{} vs.\ $h$ & 0.80 ($2.7 \times 10^{-5}$) & 0.65 (0.0019) \\
\\
\amax{} vs.\ $R_{25}$ & 0.68 ($9.2 \times 10^{-4}$) & 0.54 (0.013) \\
\lbar{} vs.\ $R_{25}$ & 0.63 (0.0028) & 0.48 (0.032) \\
\\
\amax{} vs.\ $M_{B}$ & $-0.58$ (0.0077) & $-0.54$ (0.014) \\
\lbar{} vs.\ $M_{B}$ & $-0.64$ (0.0025) & $-0.58$ (0.0079) \\
\hline
\multicolumn{3}{c}{Sb Galaxies} \\
\hline
\amax{} vs.\ $h$ & 0.67 (0.018) & 0.67 (0.017) \\
\lbar{} vs.\ $h$ & 0.67 (0.018) & 0.76 (0.004) \\
\\
\amax{} vs.\ $R_{25}$ & 0.44 (0.15) & 0.27 (0.39) \\
\lbar{} vs.\ $R_{25}$ & 0.43 (0.17) & 0.24 (0.46) \\
\\
\amax{} vs.\ $M_{B}$ & $0.01$ (0.97) & $0.06$ (0.85) \\
\lbar{} vs.\ $M_{B}$ & $0.02$ (0.95) & $0.06$ (0.86) \\
\hline
\end{tabular}

 \medskip

 As for Table~\ref{tab:correlations}, but using only those galaxies
 with measured disc scale length $h$ (see
 Section~\ref{sec:measure-discs}) and including correlations of bar
 size with $h$.
\end{table}

\subsection{Bar Sizes in Later-Type Galaxies and the Hubble Sequence}
\label{sec:hubble-seq}

\begin{table}
 \caption{Mean Bar Size (\lavg) for S0--Sd Galaxies}
 \label{tab:sizes-all}
 \begin{tabular}{@{}llll}
\hline
Types & \lavg{} [kpc] & $\lavg/R_{25}$ & $\lavg/h$ \\
\hline
All S0--Sab & $3.29 \pm 1.68$ & $0.37 \pm 0.13$ & $1.29 \pm 0.54$ \\
Sb          & $2.81 \pm 1.76$ & $0.28 \pm 0.11$ & $1.21 \pm 0.51$ \\
Sbc         & $2.35 \pm 1.22$ & $0.19 \pm 0.10$ & $0.92 \pm 0.70$ \\
Sc          & $1.60 \pm 0.96$ & $0.15 \pm 0.08$ & $0.57 \pm 0.46$ \\
Scd         & $1.15 \pm 0.55$ & $0.12 \pm 0.07$ & $0.56 \pm 0.52$ \\
Sd          & $1.30 \pm 0.50$ & $0.16 \pm 0.07$ & $0.58 \pm 0.23$ \\
All Sc--Sd  & $1.38 \pm 0.78$ & $0.14 \pm 0.07$ & $0.57 \pm 0.44$ \\
SB Sc--Sd   & $1.83 \pm 1.00$ & $0.19 \pm 0.07$ & $0.82 \pm 0.58$ \\
SAB Sc--Sd  & $1.14 \pm 0.50$ & $0.11 \pm 0.05$ & $0.40 \pm 0.22$ \\
\hline
 \end{tabular}

 \medskip
Relative and absolute sizes \lavg{} of bars for S0--Sd galaxies.  The
S0--Sab galaxies are from my sample, the Sbc--Sd galaxies are from
\nocite{m95}Martin's (1995) sample, and the Sb galaxies are from both.
For my galaxies $\lavg = $ the mean of \amax{} and \lbar, while $\lavg
=$ Martin's measurements for galaxies from his sample.

\end{table}

\begin{figure}
\includegraphics[scale=0.5]{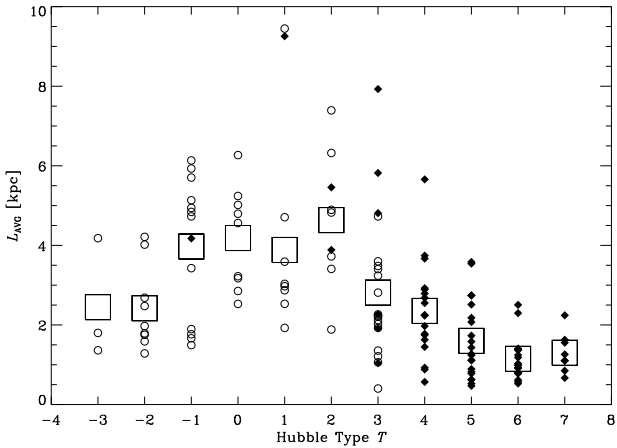}
\caption{Sizes of S0--Sdm bars in absolute terms (kpc),
combining my sample (circles) with galaxies from \citet{m95} meeting
the same selection criteria (diamonds).  All bar lengths are
deprojected; as explained in the text, I use the average of \amax{}
and \lbar{} for galaxies in my sample as the best match to Martin's
bar measurements.  Mean values for each Hubble type (using galaxies
from my sample only for S0--Sb) are indicated by the large boxes.}
\label{fig:kpc-size-all}
\end{figure}

\begin{figure}
\includegraphics[scale=0.5]{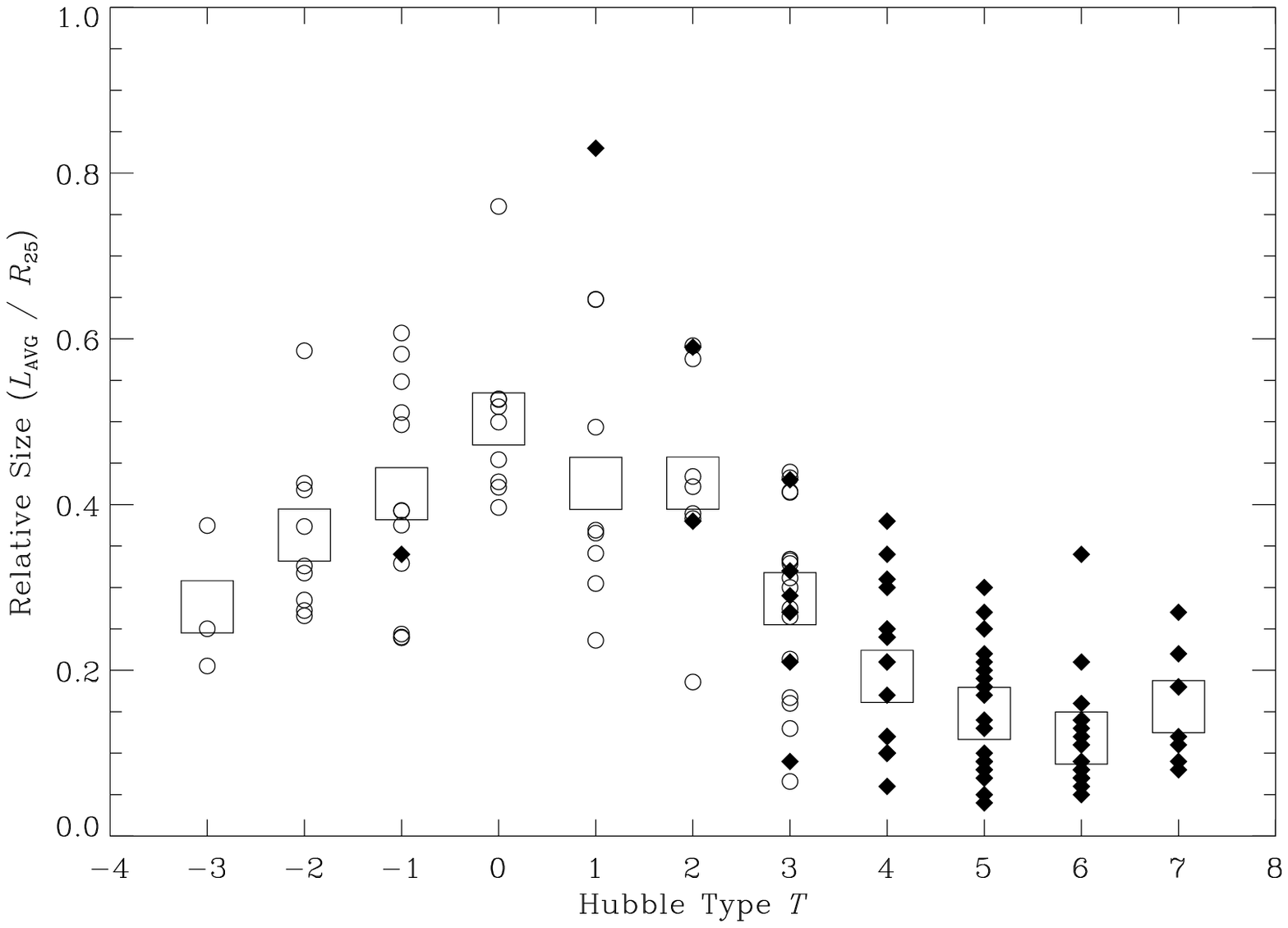}
\caption{As for Figure~\ref{fig:kpc-size-all}, but now showing size
of bars relative to disc radius $R_{25}$.}
\label{fig:r25-size-all}
\end{figure}

\begin{figure}
\includegraphics[scale=0.5]{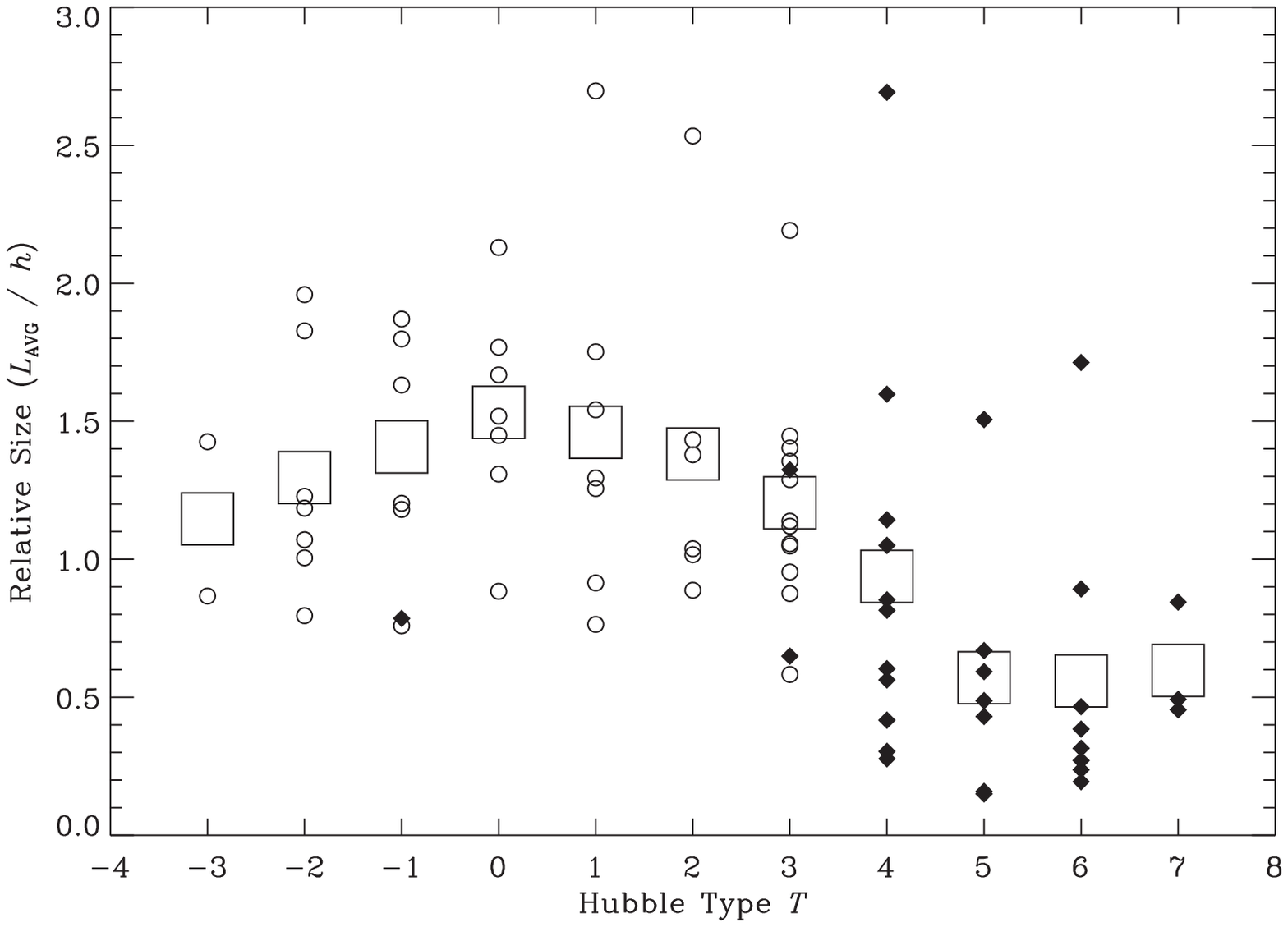}
\caption{As for Figure~\ref{fig:kpc-size-all}, but now showing size
of bars relative to the outer-disc exponential scale length $h$.}
\label{fig:h-size-all}
\end{figure}

When galaxies from the sample of \citet[][see
Section~\ref{sec:m95}]{m95} are added to the S0--Sb galaxies,
the Hubble sequence coverage extends down to Sd galaxies.
Figures~\ref{fig:kpc-size-all}--\ref{fig:h-size-all} and
Table~\ref{tab:sizes-all} show bar sizes for this combined set of
galaxies.  As noted by earlier studies \citep{ee85,m95,lsr02}, there
is a clear tendency for bars in later Hubble types to be smaller; this
is true using both relative-size measurements \textit{and} absolute
sizes.

Crudely speaking, one can divide the Hubble sequence into two zones,
plus a transition region.  Bars in S0--Sb galaxies are clearly larger
than bars in late-type (Sc--Sd) galaxies.  On average, the early-type
bars are $\approx 2.2$--2.7 times as large as late-type bars.  The K-S
probabilities that S0--Sb bars come from the same parent population as
Sc--Sd bars are $7.9 \times 10^{-11}$ for $L_{avg}$ in kpc, $3.3
\times 10^{-15}$ for $L_{avg}/R_{25}$, and $5.1 \times 10^{-6}$ for
$L_{avg}/h$; so the result is quite robust (to put it mildly).  It is
worth mentioning that \citet{m95} noted that some bars in his sample
could not be measured because ``the inner parts of the disc were
overexposed.''  This indicates a possible bias against very small bars
in Martin's final set of measurements, which apply to later-type
galaxies, so the difference could be even greater.

But where does the transition take place, and how abrupt is it?
Figures~\ref{fig:kpc-size-all} and \ref{fig:r25-size-all} make it
appear that Sb and Sbc galaxies are a continuation of later-type
galaxies: in particular, they have some very small bars ($\lavg \la 1$
kpc and $\la 0.2 R_{25}$) not found in earlier types.  However, this
continuity may be partly an illusion.  As discussed in the previous
section, the combination of large $R_{25}/h$ values for Sb galaxies
(mean = $3.9 \pm 1.0$) and diameter-limited selection leads to the
inclusion of Sb galaxies with smaller disc scale lengths than is the
case for the earlier Hubble types; since bar size correlates with
scale length for these galaxies, this leads to the inclusion of Sb
bars with smaller absolute and $R_{25}$-relative sizes.  This is
partly the case for Sbc galaxies as well: their mean $R_{25}/h$ is
$4.9 \pm 2.5$, in comparison with $4.1 \pm 1.4$ for the Sc--Sd
galaxies (there is little variation in $R_{25}/h$ among the latter
types; see Figure~\ref{fig:r25h}).

So once again the best picture is probably had by looking at bar sizes
relative to the disc scale length (Figure~\ref{fig:h-size-all}).  This
shows that Sb bars \textit{are} distinct from Sc--Sd bars, and that
they are largely indistinguishable from bars in earlier Hubble types,
as I argued in the previous section.  The transition point between the
early- and late-type regimes is thus the Sbc galaxies: their average
bar sizes are intermediate, but they include both bars as large as
those in earlier types \textit{and} bars shorter than $0.5 h$, common
in Sc--Sd galaxies but absent in the early types.

An additional, significant difference between bars in early- and
late-type galaxies is the weakness or absence of correlations between
bar size and other galaxy properties for the late-type galaxies.
Table~\ref{tab:correlations-m95} compares the various correlations
between bar size.  In all cases, the later galaxy types have weaker
bar-size correlations -- in fact, for Sc--Sd galaxies, the correlation
with $M_{B}$ is no longer statistically significant, and there is
apparently \textit{no} correlation between bar size and disc scale
length!

It is interesting -- and perhaps suspicious -- that these differences
are primarily between galaxies in my sample (Sb and earlier) and the
later types of \nocite{m95}Martin's (1995) sample.  Since the two
sample have bar (and disc scale length) measurements from different
sources, some of the dichotomy could be due to varying measurement
techniques or biases.  But at least some of it is probably real.
Table~\ref{tab:correlations-m95only} repeats the correlation analysis
using only galaxies from Martin -- including galaxies with $V > 2000$
\kms{} in order to boost the number of Sb galaxies.  The table shows
the same trends as Table~\ref{tab:correlations-m95}, including the
strong correlation of bar size with disc scale length for Sb galaxies.
This suggests that the pronounced absence of almost any correlations
in Sc--Sd galaxies between bar size and disc size or galaxy luminosity
is probably real.  As I show in the next section, this result appears
to be due, at least in part, to the fact that SB and SAB bars have
distinctive sizes in late-type galaxies.

\subsection{Bar Size and Bar Strength}

Strongly barred early-type galaxies (that is, S0--Sb galaxies with an
RC3 bar classification of SB) typically have bar sizes roughly the
same as those of early-type SAB galaxies (Table~\ref{tab:sizes}).  The
average SB bar is only $\sim10$--35\% larger than the average SAB bar
for size in kpc or relative to $R_{25}$, and 2--3\% \textit{smaller}
when size relative to disc scale length is used.  K-S tests give
probabilities of 11--94\% (depending on how the size is measured) that
the SB and SAB lengths come from the \textit{same} parent
distributions.

The parameter which \textit{does} differ significantly between SB and
SAB bars in early-type galaxies is, not surprisingly, the deprojected
ellipticity ($0.49 \pm 0.13$ for SB, $0.36 \pm 0.15$ for SAB), with a
K-S test giving only a 0.6\% probability of the same parent
distribution.  Interestingly, this is \textit{not} as true for the
\textit{observed} ellipticities: the mean ellipticity is still higher
for SB galaxies ($0.47 \pm 0.15$ versus $0.39 \pm 0.14$ for SAB), but
the K-S probability is now 22\%.

Deprojected ellipticity does correlate with bar size for S0--Sb
galaxies, but only weakly: $r_{s} = 0.46$ and 0.42 ($P = 1.2 \times
10^{-4}$ and $4.5 \times 10^{-4}$) for $\amax/R_{25}$ and \amax{} in
kpc, respectively.  These correlations are weaker when \lbar{} is used:
$r_{s} = 0.25$ and 0.29 ($P = 0.041$ and 0.021) for $\lbar/R_{25}$ and
\lbar{} in kpc, respectively.  (The correlations for bar size relative
to disc scale length are weaker still: $r_{s} = 0.09$ and $P = 0.64$
for $\lbar/h$.)  This generally agrees with \citet{chapelon99} and
\cite{lsr02}, who found very little correlation between deprojected
ellipticity or bar axis ratio and bar size for their early-type
spirals.  It is also consistent with the correlation reported by
\citet{laine02}, who used \amax{} for bar sizes -- especially since
the latter authors' samples included some Sc galaxies, for which the
correlation is stronger (see below).

In late-type galaxies, SB bars are more elliptical than SAB bars
(deprojected ellipticity $0.64 \pm 0.18$ versus $0.40 \pm 0.20$), just
as for the early-type galaxies.\footnote{This discussion omits the SBc
galaxy NGC~2835, for which \citet{m95} lists a deprojected axis ratio
of 1.0.} However, there is also a dramatic difference in bar size
between late-type strong (SB) and weak (SAB) bars.  On average, SB
bars in Sc--Sd galaxies are almost \textit{twice} the size of SAB bars
(Table~\ref{tab:sizes-all}); a K-S test shows that this difference is
significant at the 99.9\% level for bar size relative to $R_{25}$ (the
significance is 99.2\% for absolute sizes and 94\% for sizes relative
to $h$).  Since SB bars are also more elliptical than SAB bars, we
should expect a strong correlation between bar size and deprojected
ellipticity for the late-type bars, and this is indeed the case.  For
$\lavg/R_{25}$ versus deprojected ellipticity, $r_{s} = 0.73$ ($P =
9.4 \times 10^{-9}$) for Sc--Sd bars, compared with only 0.36 ($P =
0.0021$) for the S0--Sb bars.  A similar result was found by
\citet{martinet97} for a late-type (Sbc--Scd) subset of Martin's
galaxies and by \citet{chapelon99} for their ``normal'' (i.e.,
non-starbursting) Sbc and later-type galaxies.

There is some evidence for a stronger correlation between bar size and
$R_{25}$ (and perhaps also $M_{B}$) when only late-type SB bars are
considered: the Spearman correlation coefficient is 0.73 with a
probability of $P = 0.0013$ for \lavg{} versus $R_{25}$.  In contrast,
for SAB late-type galaxies $r_{s}$ is only 0.29 with $P = 0.21$.  A
similar disparity exists for correlations between relative bar size
and $M_{B}$, though they are not statistically significant for the
main sample; when all of Martin's Sc--Sd galaxies are included, the
coefficients are $-0.68$ ($P = 4.5 \times 10^{-4}$) for SB bars versus
$-0.48$ ($P = 0.0019$) for SAB bars.

Thus it appears that there may be a dichotomy between SB and SAB bars
in the late-type galaxies, with SB bars perhaps retaining some of the
characteristics (larger size, stronger correlation with $R_{25}$ and
$M_{B}$) of both SB and SAB bars in early-type galaxies.  It should be
noted that there are about twice as many SAB as SB bars in the Sc--Sd
galaxies studied here; thus the overall weakness or absence of
bar-size correlations for Sc--Sd galaxies is partly a combination of
poor correlation for the SAB bars and the dichotomy in sizes between
SB and SAB bars.  However, even when Sc--Sd bars are analyzed
independently in SB and SAB categories, there is still \textit{no}
correlation between bar size and exponential disc scale length.

\begin{table*}
\begin{minipage}{126mm}
 \caption{Correlations Between Bar Size \lavg{} and Galaxy 
 Properties for S0--Sd Galaxies}
 \label{tab:correlations-m95}
 \begin{tabular}{@{}lrllrll}
\hline
Types & $N$  &Pearson $r$ ($P$) &  Spearman $r_{s}$ ($P$) & $N$ &
Pearson $r$ ($P$) &  Spearman $r_{s}$ ($P$) \\
\hline
\multicolumn{7}{c}{Correlation with Disc Size $R_{25}$} \\
\hline
       & & \multicolumn{2}{c}{All Galaxies} & &
           \multicolumn{2}{c}{Measured $h$ Only} \\
S0--Sab & 48 & 0.76 ($4.4 \times 10^{-10}$)  & 0.75 ($1.0 \times 10^{-9}$) &
          24 & 0.72 ($7.4 \times 10^{-5}$)  & 0.66 ($4.5 \times 10^{-4}$) \\
Sb--Sbc & 40 & 0.58 ($8.3 \times 10^{-5}$)  & 0.49 (0.0013) &
          19 & 0.45 (0.051)   & 0.51 (0.025) \\
Sc--Sd  & 46 & 0.30 (0.043)   & 0.31 (0.037)  &
          18 & 0.10 (0.68)    & 0.12 (0.64) \\
\hline
\multicolumn{7}{c}{Correlation with Disc Scale Length $h$} \\
\hline
       & & \multicolumn{2}{c}{All Galaxies} & &
           \multicolumn{2}{c}{Measured $h$ Only} \\
S0--Sab &  &  &  &
          24  & 0.70 ($1.2 \times 10^{-4}$)  & 0.61 ($1.6 \times 10^{-3}$) \\
Sb--Sbc &  &  &  &
          19  & 0.32 (0.19)   & 0.34 (0.15) \\
Sc--Sd  &  &  &  &
          18  & 0.00 (1.00)   & 0.05 (0.84) \\
\hline
\multicolumn{7}{c}{Correlation with Absolute Magnitude $M_{B}$} \\
\hline
       & & \multicolumn{2}{c}{All Galaxies} & &
           \multicolumn{2}{c}{Measured $h$ Only} \\
S0--Sab & 48  & $-0.61$ ($3.7 \times 10^{-6}$)  & $-0.65$ ($6.4 \times 10^{-7}$) &
          24  & $-0.63$ (0.0011)  & $-0.62$ (0.0013) \\
Sb--Sbc & 40  & $-0.34$ (0.033)     & $-0.34$ (0.032) &
          19  & $-0.21$ (0.39)    & $-0.40$ (0.093) \\
Sc--Sd  & 46  & $-0.24$ (0.11)      & $-0.21$ (0.16)  &
          18  & $-0.06$ (0.80)  & $-0.04$ (0.89) \\
\hline
 \end{tabular}

 \medskip
 For the S0--Sb galaxies from my sample, \lavg{} is the average of
 \amax{} and \lbar; for Martin's (1995) Sb--Sd galaxies, it is his
 visual measurement $L_{b}(i)$.  For each correlation coefficient, the
 probability that the correlation is due to chance is given in
 parentheses.  The Sb--Sbc correlations include galaxies from both my
 sample and from Martin (1995).  ``Measured $h$ Only'' means those
 galaxies with measured outer disc scale lengths. $N$ is the number of
 galaxies in each subsample.
 
\end{minipage}
\end{table*}

\begin{table}
 \caption{Correlations Between Bar Size \lavg{} and Galaxy 
 Properties -- Galaxies from Martin (1995) Only}
 \label{tab:correlations-m95only}
 \begin{tabular}{@{}lrll}
\hline
Types & $N$ & Pearson $r$ &  Spearman $r_{s}$ \\
\hline
\multicolumn{4}{c}{Correlation with Disc Scale Length $h$} \\
\hline
Sb      &  7 & 0.97 ($2.1 \times 10^{-4}$)  & 0.93 (0.0025) \\
Sbc     & 16 & 0.65 (0.0064)                & 0.48 (0.060)  \\
Sc--Sd  & 24 & $-0.04$ (0.87)               & 0.02 (0.93)   \\
\hline
\multicolumn{4}{c}{Correlation with Disc Size $R_{25}$} \\
\hline
Sb      &  7 & 0.37 (0.41)                  & 0.79 (0.036) \\
Sbc     & 16 & 0.77 ($5.4 \times 10^{-4}$)  & 0.69 (0.0032) \\
Sc--Sd  & 24 & 0.09 (0.69)    & 0.08 (0.71) \\
\hline
\multicolumn{4}{c}{Correlation with Absolute Magnitude $M_{B}$} \\
\hline
Sb      &  7 & $-0.19$ (0.69)     & $-0.64$ (0.12) \\
Sbc     & 16 & $-0.65$ (0.0064)    & $-0.68$ (0.0038) \\
Sc--Sd  & 24 & $-0.03$ (0.89)     & $0.00$ (0.99)     \\
\hline
 \end{tabular}
 
 \medskip
 As for Table~\ref{tab:correlations-m95}, but using \textit{all}
 galaxies from the Martin (1995) sample with plausible disc scale
 lengths from BBA98 (except for three Virgo cluster galaxies).
\end{table}

\section{Discussion}

\subsection{Biases, Sample Incompleteness, and the Absolute Sizes of
Bars}\label{sec:bias}

All the bars studied in this paper are in galaxies classified as
barred (SB or SAB) in RC3.  Because RC3 classifications are based on
blue photographic plates, there is the possibility that some bars have
been missed, for two reasons.  First, as has been recognized for some
time, bars can be hidden in optical images due to dust and star
formation.  This is probably not a large effect, \textit{if} SAB
galaxies are included in the ``barred'' category: \citet{eskridge00}
found that the total (SB + SAB) bar fraction goes from $65 \pm 3$\%
when using $B$-band images to $73 \pm 3$\% when using $H$-band images.
However, they also found that a significant number of optically weak
(SAB) galaxies (68\%) become SB when classified in the IR, which
suggests that many optically weak bars are really ``disguised''
strong bars.

The second potential bias is the possibility that \textit{small} bars
have been missed due to resolution (and possibly saturation) effects.
Recent CCD and near-IR observations have uncovered large numbers of
small, \textit{inner} bars embedded inside large bars; most of these
went unnoticed in earlier photographic surveys
\citep[see][]{erwin04-db}.  Some galaxies classed as unbarred turn out
to have \textit{nuclear} bars small enough to have been missed in
low-resolution or nuclear-saturated photographic images
\citep[e.g.,][]{buta91,bc91,scorza98}.  So the samples may be missing
precisely those galaxies with the smallest bars.

An additional bias affects the absolute sizes (lengths in kpc).  The
samples studied here tend to exclude small, faint galaxies.  The mean
(and median) luminosity is $M_{B} = -19.3$ for the S0 galaxies,
$-19.6$ for the S0/a--Sb galaxies in my sample, and $-19.8$ for the
Sb--Sd galaxies taken from \nocite{m95}Martin's (1995) sample.  This
can be compared with the mean luminosities for cluster S0 galaxies
($M_{B} = -18.9$) and spirals ($M_{B} = -18.2$), from
\citet{jerjen97}: clearly, the bars studied here come from galaxies on
the bright ends of the distributions.  As we have seen, bar size
generally scales with disc size and with galaxy luminosity; thus,
smaller galaxies will have smaller bars.  This means that the
absolute-size distributions presented here (e.g.,
Figures~\ref{fig:kpc-size} and \ref{fig:kpc-size-all}) are biased
towards larger bars, and the mean sizes are probably overestimates for
the \textit{complete} galaxy population.

\subsection{Bar Sizes and Simulations}
\label{sec:sims}

The only reasonable way to compare the sizes of real bars with those
produced in $n$-body simulations is to use sizes relative to the disc
scale length.  In principle one can calculate sizes relative to
$R_{25}$ as well, but this requires estimating mass-to-light ratios
and the star-formation history \citep[e.g.,][]{michel-dansac04} and is
thus prone to more uncertainties.  In this section, I survey some
recent $n$-body studies in an attempt to see how well or poorly they
do at reproducing the relative sizes of real bars.  I make no attempt
to be comprehensive, and I am necessarily limited to those studies
which provide both bar sizes and some indication of disc scale length
in the region outside the bar (either as measured by the authors or
via inspection of surface density profiles).

Table~\ref{tab:n-body-sizes} summarizes results from eight different
papers.  In each case, I have included as many models from each study
as possible, though in some cases additional models are left out
because there were no bar or disc sizes for them.  One thing is
immediately apparent: \textit{simulations tend to produce large bars}.
Indeed, several simulations produce bars which are either at the upper
end of the local distribution, or are larger than any seen in nearby
galaxies.  Except for two of the earlier simulations, no $n$-body bars
are as small as typical Sc--Sd bars.  (Ironically, one of these is the
``early-type'' model CS2 of \nocite{combes93}Combes \& Elmegreen 1993,
which produced a shorter bar than their ``late-type'' model CSE.)

\citet{holley-b05} argued that when bars are triggered by satellite
interactions, rather than disc instabilities, ``the length of the bar
will depend on the mass and distance of the satellite\ldots. The
typical bar induced by this process will be much larger than those
formed through internal disk instabilities.''  Their final bar size of
$\amax/h \sim 2.6$ is indeed rather large: only two of the galaxies in
my sample and one of Martin's have bars that large
(Figures~\ref{fig:h-size} and \ref{fig:h-size-all}).  The $\amax/h
\sim 4$ bar which they report for their $B_{5}$ simulation (no
profiles shown) is clearly excessive.  None the less, the fact that
the bar size in their simulations depends on the details of the
galaxy-satellite interaction is intriguing, because it suggests that
bar sizes in real galaxies might provide clues to their host galaxies'
interaction/merger histories.  The absence of real bars with $\amax/h
> 3$ could then indicate an upper limit on past bar-forming
interactions.

Taking this argument further, one could ask if the larger bars of
early-type galaxies indicate a stronger role for interactions in their
formation.  In this vein, \citet{noguchi96} argued that bars in
early-type galaxies, with their ``flat'' major-axis profiles, are
better produced by interactions than by spontaneous disc
instabilities; the latter, he suggests, are responsible for late-type
bars.  When combined with the argument of Holley-Bockelmann et al., we
seem to get a consistent picture: the differing characteristics of
bars in early- and late-type galaxies indicate a greater role for
interactions in the evolution of early-type galaxies.  There is at
least some observational evidence for this: \citet{elmegreen90}
reported that, for early-type (Sa--Sb) galaxies, the SB fraction was
higher in binary-galaxy systems than in groups or the field; for
late-type galaxies, there was no trend with environment.  This fits
neatly into scenarios where evolution into or along the Hubble
sequence is determined primarily by the number and strength of
interactions and mergers; for example, recent simulations support the
idea that the bulges and thick discs characteristic of early-type
galaxies have grown through satellite accretion
\citep[e.g.,][]{walker96,aguerri01}.  Unfortunately, as
Table~\ref{tab:n-body-sizes} shows, both \citet{berentzen98} and
\citet{athan-m02} were able to produce extremely large bars via disc
instabilities.  Since flat bar profiles can also be produced this way
\citep[e.g.,][]{sparke87,combes93,athan-m02}, it appears that
satellite interaction may \textit{not} be a unique explanation for
early-type bars.

\citet{athan-m02} and \citet{athan03} have emphasized the importance
of angular momentum transfer in regulating the size of bars, based on
their analysis of $n$-body simulations.  Put simply, bar length is
ultimately limited by the corotation radius; if a bar slows down, the
corotation radius moves further out in the disc and the bar can grow
in length.  A bar slows if it can lose angular momentum, primarily via
resonances, to particles in the outer disc, the halo, and the bulge
(if present).  In principle, then, larger bars indicate galaxies where
the bar was able to lose more angular momentum.  This might explain
the generally large bars of early-type galaxies, since these galaxies
are more likely to have significant bulges which can act as angular
momentum sinks for the bar.  This might also explain why model B of
\citet{vk03} produces a relatively small bar, since in that model the
halo -- also an angular momentum sink -- is less massive relative to
the disc than in the A$_{1}$/A$_{2}$ models.  However, this still does
not explain why \textit{late}-type bars are as small as they are,
since even the ``bulge-less'' $n$-body simulations \citep[e.g., model
MD of ][]{athan-m02} produce bars at least a scale length in radius.

\begin{table*}
\begin{minipage}{126mm}
 \caption{Relative Bar Sizes from $n$-Body Simulations}
 \label{tab:n-body-sizes}
 \begin{tabular}{lllrrrr}
\hline
Study & Model & Type & $R_{\mathrm{bar}}$ & $h$ & $R_{\mathrm{bar}}/h$ & Notes \\
\hline
\citet{pfenniger91} &      & $n$-body     & 10   & 17  & 0.6 & \\
\citet{combes93}    & CS2  & $n$-body     &  6   & 9.4 & 0.6 & 1 \\
                    & CSE  & $n$-body     &  9.5 & 5.2 & 1.8 & 2 \\
\citet{friedli93}   & A    & $n$-body+gas &  6   & 7 & 0.9 & 3\\
\citet{berentzen98} & A    & $n$-body     &  8   & 2.2 & 3.6 & 4 \\
                    & B    & $n$-body+gas &  7   & 2.9 & 2.4 & 5 \\
\citet{athan-m02}   & MDB  & $n$-body     & 3.1--3.5 & 1.1  & 2.8--3.2 & 6 \\
                    & MD   & $n$-body     & 2.1--3.2 & 1.4  & 1.5--2.3 & 6 \\
\citet{vk03}        & A$_{1}$ & $n$-body  & 6.2--6.7 & 4.5  & 1.4--1.5 & 7 \\
                    & A$_{2}$ & $n$-body  & 4.5--5.5 & 4.2  & 1.1--1.3 & 7 \\
                    & B       & $n$-body  & 4.2--5.0 & 5.0  & 0.8--1.0 & 7 \\
\citet{holley-b05}  & $F_{5}$ & $n$-body  & 0.026    & 0.01 & 2.6 & 8 \\
\citet{immeli04}    & F    & $n$-body+gas+SF & 3--4.5 & 1.9  & 1.6--2.1 & 9 \\
\hline
\end{tabular}

\medskip

``Model'' refers to a given model from the study in question; ``Type''
indicates the type of simulation (``SF'' = star formation);
$R_{\mathrm{bar}}$ and $h$ are the bar semimajor axis and disc scale
length in model units (usually kpc).

Notes: (1) ``Early-type'' model galaxy.  (2) ``Late-type'' model
galaxy.  (3) Bar size measured at $t = 1000$ from their Fig.~2, disc
scale length for $r > 10$ kpc from their Fig.~3a.  (4) Values at $t =
65$.  (5) Values at $t = 20$--65.  (6) Range in bar sizes is their
$L_{b/a}$--$\lph$, which corresponds to my \amax--\lbar; sizes are
averages of values in their Table~1 (to match time of profiles shown
in their Fig.~5); $h$ for the MDB model is from Lia Athanassoula (private 
communication).  (7) Range in bar sizes using their two measurement
techniques.  (8) Bar size underestimates full bar length (their
Fig.~3), so probably matches \amax{} better than \lbar.  (9) Range in
bar sizes for $t = 2.8$--3.8 Gyr (their Fig.~14).

\end{minipage}
\end{table*}

\subsection{Some Implications for Secular Evolution}

In the past decade, an increasingly popular idea has been that bars
can drive long-term (``secular'') evolution of disc galaxies, perhaps
even helping to determine the present-day Hubble sequence.  The
general argument is that bars, through gas inflow and vertical
buckling, create or amplify bulges, thus shifting a galaxy from
smaller to larger bulge/disc ratio (e.g., from being an Sc galaxy to
being an Sb or Sa galaxy).  In addition, it is suggested that the
increasing central mass concentration produced by bar-driven gas
inflow can end up turning the bar \textit{into} a bulge.  This is
because a sufficiently strong central mass concentration can
apparently \textit{destroy} a bar, producing an axisymmetric,
bulgelike remnant \citep{hasan90,hasan93,norman96,berentzen98}.

A recent elaboration on the scenario of secular evolution via bar
dissolution, with specific predictions for bar \textit{sizes}, is that
of \citet{bournaud02}.  They posit multiple rounds of a sequence where
bars form, weaken or are destroyed via mass inflow, and then reform
due to gas accretion by the disc \citep[see also][]{sellwood99}.  In
their simulations, later (i.e., second, third, or even fourth!)  bars
are progressively \textit{shorter} than earlier bars.  The implication
is that early-type spirals (and S0's), whose larger bulges are built
out of multiple rounds of bar formation and bar-driven inflow, should
have smaller bars.  Unfortunately, this is clearly incompatible with
the Hubble sequence as we see it today.\footnote{At least some of the
difference may be due to the particular mode of accretion used by
\citet{bournaud02}, such that alternate accretion scenarios might
produce more realistic distributions of bar sizes (Fr\'{e}d\'{e}ric
Bournaud, private communication.)}

However, secular changes in bar size may still be relevant if we drop
the idea of bar destruction.  Most of the $n$-body bars mentioned in
the previous section have sizes measured near the end of the
simulation, and can thus be considered ``mature'' bars.  But in almost
all $n$-body simulations, bars lose angular momentum, slow down, and
increase in length as time goes by: older bars are longer than younger
bars.  The growth is usually fairly mild, but can sometimes be
dramatic -- for example, \citet{vk03} mention that the bar in their
$A_{2}$ simulation \textit{triples} in absolute length (from 1.5 to
4.5--5 kpc) between $t = 3$ and 6 Gyr.  Since the disc scale length
varies by $\la 20$\%, the \textit{relative} size of the bar also
triples, from $\sim 0.4 h$ to $\sim 1.2 h$.  This neatly spans the
range in typical relative sizes between Sc--Sd galaxies and Sa--Sb
galaxies (Table~\ref{tab:sizes-all} and Figure~\ref{fig:h-size-all}).
Could the difference in bar sizes between early- and late-type
galaxies, or the scatter in sizes for a given region of the Hubble
sequence, be at least partly a matter of bar \textit{age}?  This might
also explain the lack of a correlation between bar size and other
galaxy properties, especially disc scale length, for the late-type
galaxies (Section~\ref{sec:hubble-seq}), if disc scale length
primarily affects or determines the \textit{final} size of a bar.
Bars in Sc--Sd galaxies would then be young and/or still growing
rapidly, and thus less likely to show correlations with disc size.

This idea -- younger, shorter, and fast-growing bars in Sc--Sd
galaxies, older and longer bars in early-type galaxies -- is also
consistent with the simulations and arguments of \citet{friedli95} and
\citet{martin97}.  They combined $n$-body simulations with gas and
star formation, and noted that ``young'' bars ($< 500$ Myr old in
their simulations) had star formation concentrated along the bar major
axis, something observed in at least some SBc galaxies, while older
bars tended to have star formation confined to the nucleus, the ends
of the bars, and an inner ring/spiral surrounding the bar, a pattern
more often seen in early-type galaxies.  Based on this, they suggested
that barred Sc galaxies could evolve into barred Sb galaxies.  (The
fact that subsequent star formation is generally restricted to the
ends of the bar and an inner ring might also indicate that star
formation helps transform late-type ``exponential'' bars into
early-type ``flat'' bars, by preferentially adding stars near the ends
of the bar.)  It would be interesting to see if there is a correlation
between star-formation patterns and bar sizes in, e.g., Sbc--Sd
galaxies: are smaller bars in fact more likely to have ``young''
star-formation patterns?  There is actually a hint of this in the
sample of \citet{martin97}: they classified eleven bars into three
categories (A, B, C) based on the distribution of \hii{} regions, and
then associated these categories with increasing age, based on their
simulations.  The mean relative bar sizes of the three groups are, in
order of increasing age, $L_{avg}/R_{25} = 0.19$, 0.28, and 0.32,
which does agree with the scenario; but the sample is really too small
for this to be a valid test.

A problem with this idea is the fact that, in the few simulations
where gas has been included and bar length at different times is
reported \citep[e.g.,][]{berentzen98,immeli04}, the bar does
\textit{not} grow significantly; it may even \textit{shrink} slightly.
This may be because the bar gains angular momentum from the gas inflow
it drives, thus keeping its pattern speed constant or increasing and
keeping corotation at a small radius.  This would seem to mean that
bars in Sc--Sd galaxies, which generally have abundant gas, should not
grow significantly; this might even help explain why their bars are
generally small!  It may be that once enough of the gas within
corotation has been exhausted (e.g., converted to stars), the bars
\textit{can} grow; but whether this is possible, and whether the bars
can grow large enough, awaits further simulation.  (Both of the
studies mentioned above produced bars larger than those of typical
\textit{early}-type galaxies, so it isn't clear how relevant they are
to late-type bars.)

The alternative to these scenarios of secular evolution is that there
is something fundamentally different between early- and late-type disc
galaxies, which is reflected in their different bar properties.  It
would be fruitful to examine more carefully why some simulations
produce larger bars than others \citep[see, for example, the
discussions in][]{athan03,vk03}, and whether there are any parameters
(e.g., relative halo mass, halo structure and kinematics, and in
particular gas content) which can reliably produce the small bars of
late-type galaxies.

On a separate note, it is interesting to compare the field and Virgo
S0 galaxies in my sample.  A popular scenario for creating
\textit{cluster} S0 galaxies is ram-pressure stripping of spirals
which fall into a cluster and encounter its intracluster medium at
high speed \citep[e.g.,][]{quilis00}.  If removal of a spiral galaxy's
gas -- and subsequent aging of its stellar population without new
star formation -- is \textit{all} that happens, then we might expect
other properties, such as bar size, to remain unchanged.  Since the
most numerous bright spiral types are Sbc and Sc
\citep[e.g.,][]{eskridge02}, we should on average see smaller bars in
cluster S0's -- \textit{if} conversion of infalling spirals is the
primary formation mechanism.  However, as I showed in
Section~\ref{sec:early-sizes}, Virgo Cluster S0's tend if anything to
have \textit{larger} bars than field S0's, and their bars are
certainly consistent with the general field S0--Sb population.  This
suggests that cluster S0's are probably not just stripped and aged
spirals, or else that they are preferentially formed from
\textit{early-type} spirals.  Quilis et al.\ note that other
mechanisms may be needed to produce larger bulges and thicker discs in
stripped spirals in order to make the end products more like S0
galaxies; such mechanisms must also, it seems, ensure that cluster
S0's end up with large bars.

\subsection{Finding Bars at High Redshift}

\begin{figure}
\includegraphics[scale=0.6]{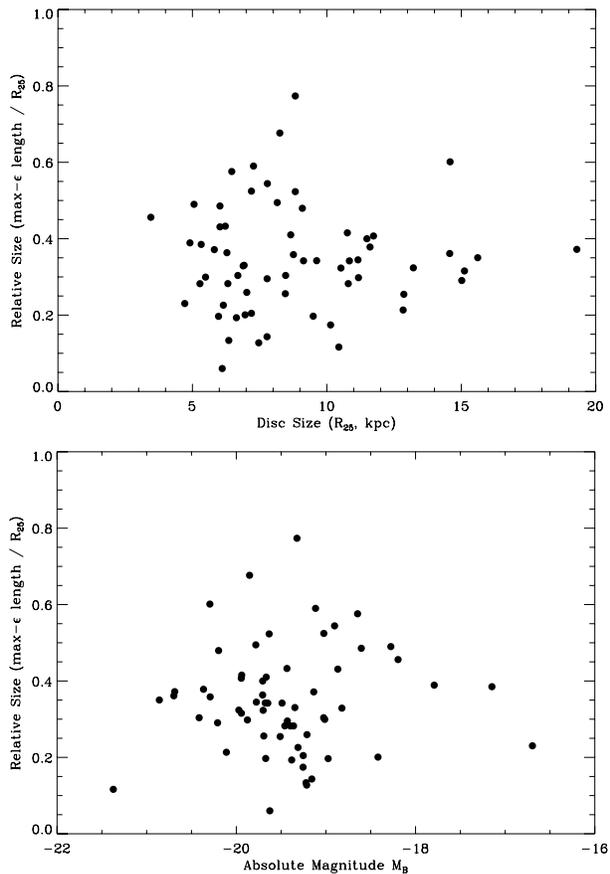}
\caption{Relative bar size $\amax / R_{25}$ versus disc size (top) and
galaxy luminosity (bottom) for S0--Sb galaxies.  Relative bar size
shows no correlation with either galaxy size or luminosity; this
suggests that galaxies smaller and/or less luminous than the sample
probably have a similar range of relative bar sizes.}
\label{fig:size-v-size}
\end{figure}

Although the samples analyzed here do not extend to very faint
magnitudes, there is no evidence for any trend in \textit{relative}
bar size with magnitude or $R_{25}$ (Figure~\ref{fig:size-v-size}).
This suggests that smaller and fainter galaxies should follow the
general bar-size--luminosity and bar-size--disc-size correlations
found here, at least for Hubble types earlier than Sc.  So if high-$z$
galaxy samples include significant numbers of faint galaxies, they
will probably include bars which are small in absolute (kpc) terms,
and thus difficult to detect.

The important point is that -- in the absence of any evolutionary
effects -- the absolute size of bars, and thus their detectability,
depends on the size, luminosity, and Hubble types of the galaxies
being studied.  Thus, a proper measure of bar fractions as a function
of redshift requires careful sample selection: the low-redshift and
high-redshift samples must contain galaxies with a similar
distribution of disc sizes or absolute magnitudes.  The latter is
probably easier to achieve, but evolutionary corrections will almost
certainly need to be applied.  Disc isophotal size will probably also
evolve with redshift, in ways perhaps less easy to model.  (Recall the
subtle bias introduced by an isophotal size limit in my sample,
leading to an Sb subsample with smaller scale lengths and absolute bar
sizes than the S0--Sab bars; see Section~\ref{sec:early-sizes}.)  The
best sample-matching might therefore be in terms of disc scale lengths
--- assuming, of course, that \textit{they} do not evolve
significantly.

Recently, \citet{vandb02} analyzed the detectability of bars and
spiral structure with redshift by artificially redshifting and
resampling $B$-band images from the Ohio State University Bright
Spiral Galaxy Survey \citep[BSGS;][]{eskridge02} to match $z = 0.7$
Hubble Deep Field (HDF) exposures.  They argued that most of the
strong bars in the BSGS galaxies would still be detectable at $z =
0.7$; the apparent absence of such bars in real HDF galaxies at $z
\sim 0.7$ \citep[e.g.,][]{vandb00} is then, they suggest, a genuine
effect.  But lurking behind this is the assumption that the BSGS is a
reasonable match to high-$z$ HDF galaxies.  As their Figure~2 shows,
the BSGS does live up to its name: the magnitude distribution peaks at
$M_{B} \approx -20.2$, and essentially all of the galaxies are
brighter than the \textit{mean} magnitude ($-18.2$) of local spirals
found by \citet{jerjen97}.  Van den Bergh et al.\ noted that deep
\textit{HST} images, such as the HDF, will sample significantly
fainter galaxies, even in the absence of luminosity evolution.  This
means that -- assuming no \textit{bar} evolution -- the average bar in
the HDF sample will probably be smaller (in kpc) than the average BSGS
bar, and therefore harder to detect \citep[see the discussion of
bar-size detectability versus redshift in][]{sheth03}.

In striking contrast to the apparent absence of optical bars at high
$z$ in the HDF, \citet{sheth03} found at least four barred galaxies
with $z \sim 1$ in the HDF by using NICMOS images, which suggests that
bandshifting effects are important.  More recent studies based on
higher-resolution ACS imaging \citep{elmegreen04,jogee04} find
comparable fractions of bars at low and high $z$; this is eloquent 
confirmation of the parallel importance of resolution, as Sheth et 
al.\ also emphasized.

\begin{figure}
\includegraphics[scale=0.5]{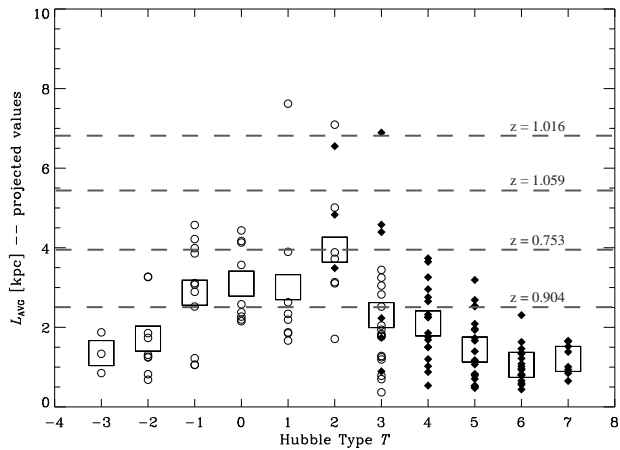}
\caption{As for Figure~\ref{fig:kpc-size-all}, but now showing
\textit{observed} (i.e., not deprojected) bar sizes.  Sizes for the
four high-$z$ bars of \citet{sheth03} are indicated by dashed
lines.}
\label{fig:high-z}
\end{figure}

Sheth et al.\ reported an average bar semimajor axis of 6 kpc for
their high-$z$, NICMOS-detected bars, and argued that this was
unusually large when compared with local galaxies -- specifically,
when compared with the BIMA Survey of Nearby Galaxies (SONG)
sample.\footnote{Note that they discuss all sizes, and plot their
ellipse fits, in terms of \textit{diameters}.} But are the bars they
found really so large?  The BIMA SONG galaxies are dominated by
intermediate and late types (21 of its 29 barred galaxies are Sbc or
later), and thus should include mostly smaller bars.  In
Figure~\ref{fig:high-z} I plot the semimajor axes of their high-$z$
bars in the context of the entire Hubble sequence.  The high-$z$ bar
sizes are based on my inspection of their NICMOS ellipse fits, with
\lavg{} = the average of \amax{} and \aten.  In general, I find
smaller values of \amax{} than they do (2.4--6.5 kpc, versus their
4--8 kpc), probably because they use the absolute peak in ellipticity,
which can be due to spiral arms outside the bar itself.  Because these
bar sizes are \textit{observed} (i.e., no deprojection was attempted),
I compare them with the observed sizes of local bars.
Figure~\ref{fig:high-z} shows that the high-$z$ bars Sheth et al.\
found are unusually large only in the context of late-type spirals;
they are large but plausible for early-type spirals.

The mean \lavg{} of the high-$z$ bars is 4.7 kpc, very close to the
NICMOS3 detection limit Sheth et al.\ suggest for $z \sim 1$.  One
might then ask, following Sheth et al., whether finding the number of
bars of that size (e.g., two with $\lavg > 4.7$ kpc) in high-$z$
spirals is at all meaningful in the context of local galaxies.  In
Figure~\ref{fig:high-z}, there are 6 out of 113 local barred spirals
with projected $\lavg > 4.7$ kpc.  This translates to 6 out of
$\sim160$ local spirals of \textit{all} bar classes -- assuming that
$\sim70$\% of local spirals with $D_{25} \geq 2.0\arcmin$ and $a/b
\leq 2.0$ have RC3 classifications of SB or SAB \citep{erwin05-rc3} --
for a local frequency of $\approx 4 \pm 2$\%.  Sheth et al.\ found
their four barred galaxies in a sample of 95 disclike galaxies with $z
> 0.7$; they noted that the total number would drop to 31 (and three
barred galaxies) if the magnitude cutoff of \citet{abraham99} was
used.  Regardless of how the parent sample is defined, the frequency
of large bars at high-$z$ ($2/95 = 2 \pm 1$\% or $2/31 = 6 \pm 4$\%)
appears consistent with the local frequency.  Of course, this analysis
assumes the parent samples are comparable, which is probably not true.
For example, the faintest of their high-$z$ barred galaxies has $H
\approx 26$, which they suggest corresponds to rest-frame $M_{B}
\approx -16$.  This is fainter than any of the galaxies in my sample,
\nocite{m95}Martin's (1995) sample, \textit{or} the BIMA SONG sample.

Finally, I note that the much larger (though currently incomplete)
sample of $z = 0.7$--1.0 disc galaxies studied by \citet{jogee04}
seems to have typical bar semimajor axes $\sim 3$ kpc, very close to
the average S0--Sb bar sizes in my sample (Table~\ref{tab:sizes}).
Again, this suggests that bar sizes at $z \sim 1$ were similar to bar
sizes in the local universe, and that high-$z$ studies will naturally
select against the small bars characteristic of late-type spirals.

\section{Summary\label{sec:summary}}

I have presented a study of bar sizes in disc galaxies, using a sample
of 65 nearby S0--Sb galaxies as well as the published bar sizes for 70
nearby Sb--Sd galaxies from \citet{m95}.  The main results are 
summarized below.

\begin{enumerate}
  \item Bars in early-type (S0--Sb) galaxies have mean absolute sizes
  (semimajor axis) of $\sim 3.3$ kpc, and mean relative sizes of $\sim
  0.38 \; R_{25}$ and $\sim 1.4 \; h$ (where $h$ is the exponential
  disc scale length).
  
  \item For these galaxies, the sizes of bars \textit{relative to disc
  scale length} is roughly constant with Hubble type.  The Sb galaxies
  in my sample appear to have smaller bars relative to $R_{25}$ in
  comparison to the S0--Sab galaxies because the Sb galaxies have, on
  average, \textit{larger} values of $R_{25}/h$.  A diameter-limited
  selection criterion then leads to smaller average scale lengths for
  these galaxies and thus bars with smaller average absolute sizes
  ($\sim 2.5$ kpc) as well.
  
  \item As has been noted earlier, bars in early-type (S0--Sb)
  galaxies are larger than those in late-type (Sc--Sd) galaxies.
  This is true regardless of how bar size is measured; bar size
  relative to disc scale length appears to be the most robust
  measurement, and the least vulnerable to selection effects.  On
  average, early-type bars are $\sim 2.5$ times larger than late-type
  bars, which have mean sizes of $\sim 1.5$ kpc, 0.14 $R_{25}$, and
  0.6 $h$.  Sbc galaxies have bars intermediate in size between the
  early and late types.
  
  \item Early-type bars show strong correlations of bar size with
  $R_{25}$ and $h$; these correlations are stronger than the known
  correlation of bar size with $M_{B}$.  But late-type bars as a whole
  show only weak correlations of bar size with $R_{25}$ and $M_{B}$,
  and \textit{no} correlation with $h$ at all.
  
  \item Strong (SB) and weak (SAB) bars in early-type galaxies differ
  primarily in ellipticity; they are very similar in size.  But
  late-type galaxies exhibit a real dichotomy: SB bars in Sc--Sd
  galaxies are on average twice the size of SAB bars, and the SB bars
  have stronger correlations of bar size with $R_{25}$ and $M_{B}$.
  
  \item Comparison with a number of recent $n$-body studies suggests
  that simulations usually produce relatively large bars (bar size
  $\ga 1.5 h$), including some bars larger than those seen in real
  galaxies.  The small bars typical of late-type galaxies (bar size
  $\sim 0.6 h$) are rare in simulations.
  
  \item Comparison with local bars shows that the recently discovered 
  $z \sim 1$ bars of \citet{sheth03} have sizes typical of those in 
  \textit{early}-type (S0--Sb) galaxies.  Because bar size scales with 
  disc size (and, less strongly, with $M_{B}$) for all but the latest 
  Hubble types, and because smaller bars are harder to detect at high 
  redshift, attempts to compare bar frequencies at different redshifts 
  must be careful to use similar samples of galaxies -- ideally 
  samples with similar disc scale lengths.
  
\end{enumerate}

\section*{Acknowledgments}

I am grateful to Paul Schechter for observations made at the MDM
Telescope, to Hans Deeg for observations made at the Isaac Newton
Telescope, and especially to Juan Carlos Vega Beltr\'an for his help
in obtaining observations at Nordic Optical Telescope; I also thank
Johan Knapen for a $K$-band image of NGC~4725.  I enjoyed helpful and
interesting conversations with a number of people, including Andrew
Cardwell, Ignacio Trujillo, Alister Graham, Alfonso L\'opez Aguerri,
Michael Pohlen, John Beckman, Victor Debattista, and Octavio
Valenzuela.  Kartik Sheth, Karin Menendez-Delmestre, Witold
Maciejewski, and Lia Athanassoula provided insightful comments on
early drafts, and Seppo Laine was quite helpful at clarifying some of
the complexities of deprojecting ellipses.  Finally, I thank the
referee, Eija Laurikainen, for a careful reading and several
suggestions that improved the paper.

This research is (partially) based on data from the ING Archive, and
on observations made with both the Isaac Newton Group of Telescopes,
operated on behalf of the UK Particle Physics and Astronomy Research
Council (PPARC) and the Nederlandse Organisatie voor Wetenschappelijk
Onderzoek (NWO) on the island of La Palma, and the Nordic Optical
Telescope, operated on the island of La Palma jointly by Denmark,
Finland, Iceland, Norway, and Sweden.  Both the ING and NOT are part
of the Spanish Observatorio del Roque de los Muchachos of the
Instituto de Astrof\'{\i}sica de Canarias.  I also used images from
the Barred and Ringed Spirals (BARS) database, for which time was
awarded by the Comit\'e Cient\'{\i}fico Internacional of the Canary
Islands Observatories.

This research made use of the NASA/IPAC Extragalactic Database (NED)
which is operated by the Jet Propulsion Laboratory, California
Institute of Technology, under contract with the National Aeronautics
and Space Administration.  It also made use of the Lyon-Meudon
Extragalactic Database (LEDA; part of HyperLeda at
http://leda.univ-lyon1.fr/).


\appendix

\section{Notes on Individual Galaxies} 

Unless otherwise noted, all disc scale lengths were measured using the
azimuthally averaged profile outside the bar region.  If no clear,
exponential profile could be determined, then no fit was performed. 
Specific exceptions, and cases where the non-exponentiality can be
traced to specific morphological features (or observational problems),
are listed below.  Galaxies which met the sample selection criteria
but which are not included in the final set of measurements are
indicated by names enclosed by brackets, or else listed at the end of 
each subsample.

\subsection{The WIYN Sample (Field S0--Sa)}

For most of these galaxies, the relevant details (including sources
for the distance measurements) are discussed in
\citet{erwin-sparke03}.  Here, I provide additional notes, primarily
on measurements of outer disc scale lengths.

\textbf{NGC 936:} Type~II outer-disc profile.

\textbf{NGC 2859:} Strong outer ring produces extreme Type~II profile.

\textbf{NGC 2880:} Outer profile is non-exponential, flattening at
large radii (probably dominated by bulge light).  The inclination is
based on the region of maximum ellipticity, where the disc appears to
dominate ($r \ap 50\arcsec$), but no clear slope can be determined.

\textbf{NGC 2962:} Type~II outer-disc profile.

\textbf{NGC 3412:} Type~II outer-disc profile.

\textbf{NGC 3489:} Bar measurements are from an unpublished WHT-INGRID
$H$-band image, due to strong dust extinction in the optical.

\textbf{NGC 3729:} The outer-disc scale length and revised outer-disc
orientation are from a Sloan $r$-band image obtained with the INT-WFC
(Erwin, Pohlen, \& Beckman 2005, in prep), since the WIYN images were
taken during full moon.

\textbf{NGC 3945:} Strong outer ring produces extreme Type~II profile;
the inclination has been updated using a high-quality $r$-band image
from the Wide Field Camera of the 2.5m Isaac Newton Telescope
(INT-WFC, La Palma; Erwin, Pohlen, \& Beckman 2005, in prep).

\textbf{NGC 4143:} The outer-disc scale length is from a Sloan
$r$-band image obtained with the INT-WFC (Erwin, Pohlen, \& Beckman
2005, in prep).

\textbf{NGC 4203:} Type~II outer-disc profile.  The length \lbar{} of
the bar, based on the ellipticity minimum, is undoubtedly an
overestimate; since this galaxy is nearly face-on and lacks spiral
arms, the ellipse-fit measurements $\amin$ and $\aten$ are misleading
or undefined.

\textbf{NGC 4245:} Outer disc orientation and scale-length
measurements are from a Sloan $r$-band image obtained with the INT-WFC
(Erwin, Pohlen, \& Beckman 2005, in prep).

\textbf{NGC 4665:} Type~II outer-disc profile.

The following galaxies in the WIYN Sample were eliminated because they
appeared to lack bars, or because they were too dusty and highly
inclined for accurate measurements of their bars \citep[for details,
see][]{erwin-sparke03}: NGC~2655, 2685, 3032, and 4310.

\subsection{Virgo S0}

\textbf{NGC 4267:} All measurements are from Nordic Optical Telescope
(NOT) $R$-band images; bar measurements agree very well with the
$H$-band measurements of \citet{jungwiert97}.

\textbf{NGC 4340:} All measurements are from MDM $R$-band images (bar 
measurements agree very well with $J$- and $K$-band measurements from 
BARS Project images), except for the outer-disc scale length, which 
is from the $R$-band image of \citet{frei96}.

\textbf{NGC 4371:} Bar measurements are from WIYN $R$-band images
measurements, but the outer-disc inclination and scale length are from
deeper INT-WFC $r$-band images (Erwin, Pohlen, \& Beckman 2005, in
prep).

\textbf{[NGC 4435]:} Since this galaxy is apparently interacting with
its neighbor NGC~4438, and possibly edge-on as well 
\citep[e.g.,][]{kenney95}, I excluded it from the sample.

\textbf{NGC 4477:} All measurements are from the $R$-band image of
\citet{frei96}.

\textbf{[NGC 4531]:} This galaxy has a dusty inner spiral, but no
evidence for a bar, despite its SB0 classification.

\textbf{NGC 4596:} All measurements are from BARS Project $R$-band
images (taken with the Prime Focus Camera of the Isaac Newton
Telescope), except for the outer-disc orientation and inclination,
which are from a deeper $I$-band image.

\textbf{NGC 4608:} All measurements are from NOT $R$-band images.

\textbf{NGC 4612:} All measurements are from MDM $R$-band images;
these agree well with measurements made using the $R$-band image of
\citet{frei96}.

\textbf{NGC 4754:} All measurements from WIYN $R$-band images, except 
that the outer-disc scale length was determined from the $R$-band 
image of \citet{frei96}, due to strong background variations in the 
WIYN image.

\subsection{Field Sab--Sb}

Unless otherwise noted, bar and disc measurements for these galaxies
were made using $R$-band images from the Nordic Optical Telescope
(NOT), supplemented in some cases by $J$ and $K_{s}$ images from the 
William Herschel Telescope (WHT).

\textbf{[NGC 278]:} Both optical and near-IR images indicate that this
SAB galaxy is not actually barred \citep[e.g.,][]{eskridge00}.

\textbf{[NGC 2146]:} This galaxy is severely distorted and almost
certainly interacting; near-IR images suggest there is probably no
bar.

\textbf{NGC 2712:} Bar measurements are from near-IR images, due to
strong dust extinction in the $R$-band.  The outer-disc scale length
is from an archival $R$-band INT-WFC image; disc orientation is from
\hi{} kinematics \citep{krumm82}.
 
\textbf{NGC 3351:} Bar measurements are from the $r$-band image of
\citet{frei96}; the outer-disc profile is Type~II. Distance is from
\textit{HST} Cepheid measurements \citep{freedman01}.

\textbf{NGC 3368:} Because the (outer) bar is very dusty in the
optical, measurements were made using the $K$-band image of
\citet{mh2001}.  The outer-disc PA and inclination are from WIYN
$R$-band images, which agree well with measurements made using the
\citet{frei96} $R$-band image and with kinematic line-of-nodes from
both the \hi{} study of \citet{schneider89}, as quoted in
\citet{sakamoto99}, and the near-nuclear 2D spectroscopy of
\citet{silchenko03}.  Type~II outer-disc profile.

\textbf{[NGC 3455]:} Inspection of $R$-band and NICMOS2 F160W images 
strongly suggests that this SAB galaxy is not actually barred.

\textbf{NGC 3504:} Bar measurements are from a BARS Project $I$-band
image from the NOT (no $R$-band images are available).  Outer disc
orientation is from \citet{grosbol85} and \citet{kenney93}.  Although
the inclination is uncertain, deprojection is not a major issue given
that the bar is almost aligned with the outer-disc major axis.

\textbf{NGC 3982:} Type~II outer-disc profile.  Bar measurements are
from a NICMOS2 F160W image; outer-disc PA is from
\citet{sanchez-portal00}.

\textbf{NGC 4102:} Type~II outer-disc profile.  Bar measurements are
from a NICMOS3 F160W image.

\textbf{NGC 4151:} Bar measurements and outer-disc scale length are
from a BARS Project $R$-band image (taken with the INT Prime Focus
Camera).  Outer disc orientation and inclination from the \hi{}
kinematics \citep{bosma77,pedlar92}.

\textbf{NGC 4319:} Bar measurements are from archival Jacob Kapteyn
Telescope $R$-band images, obtained from the Isaac Newton Group
Archive, and from an unpublished WHT $J$-band image.  Outer disc
orientation and inclination is from \citet{grosbol85}.

\textbf{NGC 4725:} The large-scale bar in this galaxy is peculiar and
somewhat difficult to measure, since it twists sharply with radius (it
is similar to NGC~3185 and NGC~5377 in this respect).  None the less,
there is a clear ellipticity maximum very close to the inner ring
(which itself defines \lbar); this agrees fairly well with the
measurements of \citet{m95} and \citet{chapelon99}.  Although the
galaxy is somewhat dusty, the $R$-band bar measurements agree very
well with measurements made with a $K$-band image kindly provided by
Johan Knapen.  The outer disc scale length is from the $r$-band image
of \citet{frei96}; distance is from \textit{HST} Cepheid measurements
\citep{freedman01}.  The ellipticity of the outer disc is uncertain,
due to the presence of two strong spiral arms, so the inclination is
based in inverting the Tully-Fisher relation using the $H$-band
magnitude of \citet{gavazzi96}, the \hi{} width $W_{20}$ from
\nocite{rc3}RC3, the Cepheid distance, and the $H$-band Tully-Fisher
relation as given in \citet[][p.~425]{bm98}.

\textbf{[NGC 4941]:} \citet{greusard00} argued that this galaxy has a
nuclear bar but no large-scale bar, based on their near-IR images; on
the other hand, \citet{eskridge02} classified it as SAB using
lower-resolution $H$-band images, so it is not clear whether this is a
single- or double-barred galaxy.

\textbf{NGC 5740:} Bar measurements are from near-IR images, due to
strong dust extinction in the $R$-band.  The outer-disc scale length,
from fits to $r < 100\arcsec$ (this is another Type~III profile, so
the disc beyond that radius has a shallower slope) agrees beautifully
with \nocite{courteau96b}Courteau's (1996) $r$-band $h = 18.3\arcsec$.

\textbf{NGC 5806:} The bar presence is uncertain, at least in the
$R$-band, though there is a clear ellipticity peak.  The outer
surface-brightness profile is Type~III; the disc scale length is from
the extended exponential region ($r \ap 55$--100\arcsec) outside the
bar.  This scale length matches the $r$-band scale length
(28.8\arcsec) of \citet{courteau96b} quite well, though not the
$V$-band major-axis scale length (15\arcsec) of \citet{bba98}.

\textbf{NGC 5832:} \amax{} is taken from a minimum in the PA.

\textbf{NGC 5957:} \amax{} is taken from a maximum in the PA.

\textbf{NGC 6012:} There are several ellipticity maxima in the ellipse
fits within the bar; in this case, $\amax$ is taken from the extremal
value of the position angle.  The outer disc orientation is from the
$R$-band image; since the outer ellipticity is uncertain, the
inclination is based on inverting the Tully-Fisher relation, using the
$H$-band magnitude of \citet{dejong94}, the \hi{} width $W_{20}$ from
\nocite{rc3}RC3, the LEDA distance, and the $H$-band Tully-Fisher
relation as given in \nocite{bm98}(Binney \& Merrifield 1998, p.~425).
The outer-disc scale length was measured from an INT-WFC $r$-band 
image (Erwin, Pohlen, \& Beckman 2005, in prep).

\textbf{NGC 7177:} Bar and disc measurements are from an archival
INT-WFC $R$-band image.  The outer-disc scale length is from the $r =
35$--60\arcsec{} region, and agrees very well with measurements by
\citet{dejong94} and \citet{graham01}.  Outside this region, the
profile flattens; this is another Type~III profile.

\textbf{UGC 3685:} The outer-disc scale length was measured from an
INT-WFC $r$-band image (Erwin, Pohlen, \& Beckman 2005, in prep).  The
outer-disc inclination is from \citet{kornreich98}, with PA from \hi{}
kinematics \citet{kornreich00}.

\subsection{Data for Galaxies from Martin (1995)}
\label{app:m95}

For the galaxies of Martin (1995), I took total blue magnitudes
($B_{tc}$) from LEDA. Distances are mostly from LEDA as well, using
the velocities corrected for Virgo-centric motion and $H_{0} = 75$
\kms{} Mpc$^{-1}$, except for cases where more accurate distances were
available.  These were mainly \textit{HST} Cepheid distances from
\citet{freedman01}, which I used for NGC~925, 1365, 3198, and 5457.
In two cases, I used distances to galaxies in the same group as one of
Martin's galaxies: NGC~4236 is in the same group as NGC~2403 (M81),
which has an \textit{HST} Cepheid distance, while NGC~5236 is in the
same group as NGC~5253 \citep[$D = 4.2$ Mpc from Cepheids;][]{saha95}
and NGC~5128 \citep[$D = 4.2$ Mpc from surface-brightness
fluctuations;][]{tonry01}.

Some of the galaxies in Table~1 of Martin (1995) have incorrect
numerical Hubble types ($T$) listed, though the full RC3 types are
correct: NGC~1156, 1288, 1433, 3614, 4214, 4304, 5350, and IC~1953.
Finally, ``NGC~4891'' is really NGC~4397, and ``New1'' is
MCG-01-03-085 (also listed as ``Shapley-Ames 1'' in NED).

\end{document}